\documentclass[aip,adv,reprint,onecolumn,noshowkeys,superscriptaddress]{revtex4-2}
\usepackage{graphicx,dcolumn,bm,xcolor,multirow,amscd,amsmath,amssymb,amsfonts,physics,longtable,wrapfig,siunitx,longtable}\usepackage[version=4]{mhchem}
\usepackage{color,soul}

\usepackage[utf8]{inputenc}  
\usepackage[T1]{fontenc}
\usepackage{listings}
\usepackage{listingsutf8}
\usepackage{simplewick}
\usepackage{algorithm2e}
\usepackage{float}
\usepackage[most]{tcolorbox}
\newtcblisting{commandshell}{colback=black,colupper=white,colframe=yellow!75!black,
listing only,listing options={style=tcblatex,language=sh},
every listing line={\textcolor{red}{\small\ttfamily\bfseries DeathStar \$> }}}
\definecolor{identifiercolor}{rgb}{.1,.1,.1}
\definecolor{stringcolor}{gray}{0.15}
\definecolor{inactivecolor}{rgb}{0.15,0.15,0.5}
\newcommand{\red}[1]{\textcolor{red}{#1}}
\newcommand{\blue}[1]{\textcolor{blue}{#1}}
\newcommand{\purple}[1]{\textcolor{violet}{#1}}
\definecolor{darkgreen}{RGB}{0, 180, 0}
\newcommand{\green}[1]{\textcolor{darkgreen}{#1}}
\newcommand{\brown}[1]{\textcolor{magenta}{#1}}

\usepackage{tikz}
\usetikzlibrary{shapes.geometric,arrows,backgrounds,fit}
\usetikzlibrary{arrows.meta}
\definecolor{ballblue}{rgb}{0.37, 0.67, 1.0}
\definecolor{aliceblue}{rgb}{0.54, 0.97, 1.0}
\definecolor{airforceblue}{rgb}{0.26, 0.80, 0.750}
\tikzstyle{startstop} = [rectangle, rounded corners, minimum width=3cm, minimum height=1cm,text centered, draw=black, fill=red!30]
\tikzstyle{io} = [rectangle, minimum width=3cm, minimum height=1cm, text centered, draw=black, fill=airforceblue!30]
\tikzstyle{process} = [rectangle, minimum width=3cm ,minimum height=1.2cm, draw=black, fill=ballblue!30]
\tikzstyle{subalgorithm} = [rectangle, minimum width=3cm ,minimum height=1.2cm, draw=black, fill= ballblue!60 ]
\tikzstyle{decision} = [diamond, minimum width=3cm, minimum height=1cm, text centered, draw=black, fill=green!30]
\tikzstyle{arrow} = [thick,->,>=stealth]

\usepackage{hyperref}
\hypersetup{
    colorlinks,
    linkcolor={red!50!black},
    citecolor={red!70!black},
    urlcolor={red!80!black}
}

\DeclareUnicodeCharacter{2212}{-}
\DeclareUnicodeCharacter{2009}{\,} 

\definecolor{codegreen}{rgb}{0.58,0.4,0.2}
\definecolor{codegray}{rgb}{0.5,0.5,0.5}
\definecolor{codepurple}{rgb}{0.25,0.35,0.55}
\definecolor{codeblue}{rgb}{0.30,0.60,0.8}
\definecolor{backcolour}{rgb}{0.98,0.98,0.98}
\definecolor{mygray}{rgb}{0.5,0.5,0.5}

\definecolor{sqred}{rgb}{0.85,0.1,0.1}
\definecolor{sqgreen}{rgb}{0.25,0.65,0.15}
\definecolor{sqorange}{rgb}{0.90,0.50,0.15}
\definecolor{sqblue}{rgb}{0.10,0.3,0.60}

\lstdefinestyle{mystyle}{
    backgroundcolor=\color{backcolour},
    commentstyle=\color{codegreen},
    keywordstyle=\color{codeblue},
    numberstyle=\tiny\color{codegray},
    stringstyle=\color{codepurple},
    basicstyle=\ttfamily\footnotesize,
    breakatwhitespace=false,
    breaklines=true,
    captionpos=b,
    keepspaces=true,
    numbers=left,
    numbersep=5pt,
    numberstyle=\ttfamily\tiny\color{mygray},
    showspaces=false,
    showstringspaces=false,
    showtabs=false,
    tabsize=2
}

\newcommand{\ie}{\textit{i.e.}}

\newcommand{\mat}{\textsc{mathematica}\xspace}
\newcommand{\alert}[1]{\textcolor{black}{#1}}
\usepackage[normalem]{ulem}

\newcommand{\hI}{\Hat{1}}
\newcommand{\hR}{\Hat{\mathcal{R}}}
\newcommand{\hL}{\Hat{\mathcal{L}}}
\newcommand{\hH}{\Hat{H}}
\newcommand{\hHN}{\hH_{\text{N}}}
\newcommand{\hFN}{\Hat{F}_{\text{N}}}
\newcommand{\hVN}{\Hat{V}_{\text{N}}}

\newcommand{\bH}{\Bar{H}}
\newcommand{\bHN}{\Bar{H}_{\text{N}}}

\newcommand{\hT}{\Hat{T}}

\newcommand{\hA}{\Hat{A}}
\newcommand{\hB}{\Hat{B}}
\newcommand{\hC}{\Hat{C}}

\newcommand{\cre}[1]{\Hat{#1}^{\dag}}
\newcommand{\ani}[1]{\Hat{#1}}
\newcommand{\crn}[2]{\Hat{#1}^{\dag}_{#2}}
\newcommand{\ann}[2]{\Hat{#1}_{#2}}

\newcommand{\Ph}[2]{\Phi_{#1}^{#2}}
\newcommand{\Ps}[2]{\Psi_{#1}^{#2}}

\newcommand{\no}[1]{ \{ #1 \}}
\newcommand{\co}[1]{\left(#1 \right)_{C}}

\newcommand{\ar}[2]{r_{#1}^{#2}}
\newcommand{\at}[2]{t_{#1}^{#2}}
\newcommand{\al}[2]{l_{#1}^{#2}}

\newcommand{\mbt}[2]{\chi_{#1}^{#2}}
\newcommand{\eriel}[2]{v_{#1}^{#2}}
\newcommand{\fockel}[2]{f_{#1}^{#2}}

\newcommand{\bEOM}{\boldsymbol{\mathcal{\bar{H}}}}

\newcommand{\ind}[2]{{#1}_{#2}}

\newcommand{\LCPQ}{Laboratoire de Chimie et Physique Quantiques (UMR 5626), Universit\'e de Toulouse, CNRS, UPS, France}

\begin{document}	

\title{Equation Generator for Equation-of-Motion Coupled Cluster Assisted by Computer Algebra System}

\author{Ra\'ul \surname{Quintero-Monsebaiz}}
	\email{rquintero@irsamc.ups-tlse.fr}
	\affiliation{\LCPQ}
\author{Pierre-Fran\c{c}ois \surname{Loos}}
	\email{loos@irsamc.ups-tlse.fr}
	\affiliation{\LCPQ}

\begin{abstract}
We present an equation generator algorithm that utilizes second-quantized operators in normal order with respect to a correlated or non-correlated reference and the corresponding Wick theorem. The algorithm proposed here, written with \textsc{mathematica}, enables the generation of non-redundant strings of second-quantized operators that, after classification, are directly assigned to many-body term quantities used to construct the many-body Hamiltonian. We demonstrate the capabilities of the algorithm by computing the CC amplitude equations and various blocks of the equation-of-motion many-body Hamiltonian. A comprehensive description of this four-step algorithm is provided alongside concrete examples.
\end{abstract}

\maketitle

\section{Introduction}

The increase in computational capabilities has led us to explore more complex wave function ans\"atze. As these ans\"atze become more complicated, deriving the corresponding algebraic equations also becomes more tedious, time-consuming, and, more importantly, subject to human errors. An important feature of many-body quantum chemistry methods is that most of them can be written in the second quantization formalism based on established commutation or anti-commutation relations between second-quantized operators. \cite{dirac1927} In the second quantization formalism, first introduced in quantum field theory, the wave function and the operators are expressed as products of creation and annihilation operators, also known as strings. These strings can be evaluated using either \alert{(anti-)commutation} rules, diagrammatic techniques, \cite{shavitt2009} or Wick's theorem.\cite{wick1950} \alert{The diagrammatic approach proves to be more efficient since it avoids redundant terms.}

 These rules can then be programmed with the aid of symbolic algebra software, allowing us to write chains of products of these operators as tensors that can be efficiently evaluated by modern tensor contraction tools. \cite{matthews2018,abdelfattah2016,paul2017,justus2015,solomonik2014,herault2020,psarras2021}

In the context of coupled-cluster (CC) methods, \cite{coester1957,coester1960,cizek1966,crawford2000,shavitt2009} various equation generators have emerged to automate the process of deriving and implementing the corresponding equations, replacing the manual derivation and error-prone implementation in electronic structure software. In their pioneering work, Janssen and Schaefer generated and automatically implemented the open-shell CC equations by utilizing second-quantized strings. \cite{janssen1991} In a similar vein, Li and Paldus automated the implementation of spin-adapted open-shell CC equations, with the added benefit of utilizing the unitary group formalism that allows for the efficient handling of low-spin states. \cite{li1994} In 2001, K\'allay and Surj\'an proposed a general-order CC code by combining diagrammatic many-body perturbation theory and string-based configuration interaction. \cite{kallay2001} In such a way, the CC equations were written in terms of diagrams and stored as strings. This technique was then generalized to state-specific multireference CC, \cite{kallay2010} excited states computed within the linear response formalism,\cite{kallay2004} and approximate treatment of higher excitations. \cite{kallay2005}

Adopting the same design philosophy as Janssen and Schaefer, Hirata implemented the Tensor Contraction Engine (\textsc{tce}) \cite{hirata2003,hirata2004,alexander2006,hirata2006} that performs the manipulation of second-quantized operators and the generation of the computer code. The main distinction is that \textsc{tce} takes advantage of spin, spatial, and index permutation symmetries at every stage of the calculations, reducing the computational cost and storage requirement. Later on, Hirata also developed an equation generator for equation-of-motion (EOM) CC \cite{stanton1993,barlett2012,krylov2008,musial2020,emrich1981,sekino1984,comeau1993} for neutral excitations (EE-EOM-CC), \cite{stanton1993,barlett2012,krylov2008,musial2020} ionization potentials (IP-EOM-CC),\cite{stanton1994,bartlett1997,stanton1999,muneaki2006,bomble2005,stanton1999} and electron affinities (EA-EOM-CC). \cite{coester1957,nooijen1995a,nooijen1995,muneaki2007} Hanrath \textit{et al.}~proposed an improved version of \textsc{tce} by implementing the matrix-matrix multiplication-based antisymmetric tensor contraction for general CC. \cite{hanrath2010,engels2011} Meanwhile, Kong \textit{et al.}~developed the EOM version of state-specific multireference CC together with the automated implementation of such complicated equations.\cite{kong2009} This was later generalized to arbitrary order. \cite{das2010,evangelista2011,hanauer2011,hanauer2012} Likewise, Shiozaki implemented explicitly-correlated versions of CC in a similar way. \cite{shiozaki2008,kohn2008} 

More recently appeared (i) the Symbolic Manipulation Interpreter for Theoretical cHemistry (\textsc{smith3})\cite{matthew2015} for complete active space methods that implies partial contractions of second-quantized operators, (ii) \texttt{p$^{\dagger}$q} developed by Rubin and DePrince III \cite{rubin2021} that combines \texttt{C++} and \texttt{python} for proof-of-concept implementation of many-body quantum chemistry methods, and (iii) \textsc{wick\&d} \cite{evangelista2022} that presents a strategy to evaluate Mukherjee's \cite{mukherjee1997} and Kutzelnigg's \cite{kutzelnigg1997} version of Wick's theorem in the case of an arbitrary number of orbital subspaces. Each of the above programs is a clear example of the progress that has been achieved in the last three decades with respect to equation generators in quantum chemistry.

The aim of the present paper is to describe a four-step algorithm based on Wick's theorem to obtain the working equations of EE-EOM-CC, IP-EOM-CC, EA-EOM-CC, as well as other quantities of interest.
In particular, the direct way of obtaining intermediates facilitates the comparison between many-body perturbation theory and CC methods. \cite{monino2022,berkelbach2018,scuseria2008,scuseria2013,lange2018} 
The present algorithm is implemented with the computer algebra system, \mat, \cite{mathematica} in order to be able to generate and manipulate the equations in a user-friendly way.

This article is organized as follows. Section \ref{sec:theo} gathers all theoretical details.
In particular, Sec.~\ref{subsec:normalorder} describes Wick's theorem and the concept of normal ordering. 
Then, in Secs.~\ref{subsec:cc} and \ref{subsec:eomcc}, we report the main theoretical details behind the CC and EOM-CC equations, respectively.
Finally, in Sec.~\ref{sec:eomcceqgen}, we provide a detailed description of our algorithm. 
Our conclusions are drawn in Sec.~\ref{sec:conclusion}.

\section{Theory}
\label{sec:theo}

\subsection{Normal ordering and Wick's theorem}
\label{subsec:normalorder}

Our goal is to derive automatically the EOM-CC working equations which is a matrix eigenvalue problem. 
Each element of the EOM-CC effective Hamiltonian is an expectation value formed by products of operators.
(We shall discuss this point in more detail later in this section). 
An efficient way to evaluate these products of operators is via their second-quantized form. \cite{helgaker2013,surjan2012,szabo2012} 

As a starting point, let us introduce the reference state which is represented as a single determinant
\begin{equation}
	\ket*{0} =\ket*{ijk \cdots} = \cre{i} \cre{j} \cre{k} \cdots \ket*{}.
\label{eq:Fermivacum}
\end{equation}
obtained by acting on the physical (or true) vacuum $\ket*{}$ with creation operators $\cre{i}$, $\cre{j}$, $\cre{k}$, $\ldots$.
This reference state $\ket*{0}$, typically taken as the Hartree-Fock (HF) ground-state determinant, can be used as an alternative vacuum, named Fermi vacuum. \cite{crawford2000,shavitt2009} 
One property of the Fermi vacuum is that one can redefine the creation operators relative to it as particles ($\cre{a}$, $\cre{b}$, $\cre{c}$, $\ldots$) and holes ($\ani{i}$, $\ani{j}$, $\ani{k}$, $\ldots$).\cite{bogoliubob1959,paldus1975}
The particle creation operators create particle states above the Fermi level while the hole creation operators remove particle states below the Fermi level (particle-hole formalism). 
Because one cannot annihilate a hole or remove a particle in the Fermi vacuum, we have $\cre{i}\ket*{0} = 0$ and $\ani{a}\ket*{0} = 0$.

It is also possible to define ``neutral'' excited determinants by operating the same number of particle and hole operators onto the Fermi vacuum:
\begin{equation}
	\cre{a}\cre{b} \cre{c}\cdots\ani{k}\ani{j}\ani{i}   \ket*{0} = \ket*{\Ph{ijk\cdots}{abc\cdots}}.
\label{eq:excdet}
\end{equation}
In such a way, it is also possible to define ``charged'' excited determinants when the number of particles and holes are different. 
For example, ionized and electron-attached determinants can be represented respectively, as follows

\begin{equation}
\begin{aligned}
	\cre{a}\ani{j}\ani{i} \ket{0} &= \ket*{\Ph{ij}{a}}, \\
	\cre{a}\cre{b} \ani{i}\ket{0} &= \ket*{\Ph{i}{ab}}.
\end{aligned}
\end{equation}
To preserve the antisymmetry of the electronic wave function, the second-quantized fermionic operators fulfill anti-commutation rules
\begin{equation}
\begin{aligned}
\label{eq:anticommrules}
	\ani{p} \ani{q} + \ani{q} \ani{p} & = 0,
	\\
	\cre{p} \cre{q} + \cre{q} \cre{p} & =0, 
	\\
	\cre{p} \ani{q} + \cre{q} \ani{p} & = \delta_{pq},
\end{aligned}
\end{equation}
where $\ani{p}$, $\ani{q}$, $\ani{r}$, $\ldots$ are arbitrary operators that can be either hole or particle operators, and $\delta_{pq}$ is a Kronecker delta.

Besides the anti-commutation rules defined in Eq.~\eqref{eq:anticommrules}, there is another powerful ``bookkeeping system'' called normal ordering, \cite{merzbacher1998} that consists of placing all the creation operators to the left and all annihilation operators to the right. 
Therefore, applying the Fermi vacuum on a normal-ordered string yield zero.
At this stage, it is convenient to introduce a more compact notation to take full advantage of this bookkeeping system. To do so, we define the contraction of arbitrary operators
\begin{equation}
\contraction{}{\cre{p}}{}{ \ani{q}} \cre{p} \ani{q} 
	= \cre{p} \ani{q}  - \no{\cre{p} \ani{q}}
	= \cre{p} \ani{q}  + \ani{q}\cre{p} 
	=\delta_{pq},
\label{eq:nofermy}
\end{equation}
where $\no{\cre{p} \ani{q}}$ means that the product $\cre{p} \ani{q}$ is normal ordered with respect the Fermi vacuum. 
By definition, the only non-zero contractions in Eq.~\eqref{eq:nofermy} are
\begin{equation}
\begin{aligned}
	\contraction{}{\hat{a}}{}{\hat{b}} \hat{a} \hat{b}^{\dagger} & = \delta_{ab},
	\\
	\contraction{}{\hat{i}}{^{\dagger} }{\hat{j}} \hat{i}^{\dagger} \hat{j} & =\delta_{ij}.
\label{eq:nonzerocont}
\end{aligned}
\end{equation}
When a product of second-quantized operators is normal-ordered, we name it a normal product.

We are now in a position to introduce the generalized Wick theorem stating that a product of creation and annihilation operators is equal to their normal product plus the sum of all possible contractions. \cite{shavitt2009} 
For example, applying this theorem to a given string of normal-ordered operators (which is typically what one has to do to derive the EOM-CC equations), one gets
\begin{equation}
    \no{\hA}\no{\hB}\no{\hC}\cdots=\no{\hA\hB\hC\cdots}
    +
     \sum_{\text{singles}}
    \no{ 
    \contraction        {}    {\hA} {\hB\hC} {\cdots }
    \hA \hB  \hC \cdots  \cdots
    }
    +
    \sum_{\text{doubles}}
    \no{ 
    \contraction        {}    {\hA} {\hB\hC} {\cdots }
    \contraction[2ex] {\hA}   {\hB}   { \hC\hC}  {\cdots }
    \hA \hB  \hC \cdots  \cdots
    }
    +
    \cdots
    +
    \sum_{\text{fully contracted}}
    \no{ 
    \contraction        {}    {\hA} {\hB\hC} {\cdots }
    \contraction[2ex] {\hA}   {\hB}   { \hC\hC}  {\cdots }
    \contraction[3ex] {\hA\hB}{\hC}  {\cdots}    {\cdots} 
    \hA \hB  \hC \cdots  \cdots
    },
\label{eq:wick}
\end{equation}
where the first term on the right-hand side is the normal product, the second term contains all the single contractions, the third contains all double contractions, and the last sum gathers the so-called fully contracted terms.
The key feature of this theorem is that, if we evaluate the Fermi vacuum expectation value associated with Eq.~\eqref{eq:wick}, the only non-zero terms correspond to the fully contracted ones, \ie,
\begin{equation}
    \mel{0}{\no{\hA}\no{\hB}\no{\hC}\cdots}{0}
    =
    \sum_{\text{fully contracted}}
    \mel{0}{
    \no{ 
    \contraction        {}    {\hA} {\hB\hC} {\cdots }
    \contraction[2ex] {\hA}   {\hB}   { \hC\hC}  {\cdots }
    \contraction[3ex] {\hA\hB}{\hC}  {\cdots}    {\cdots} 
    \hA \hB  \hC \cdots  \cdots
    }
    }{0}
\label{eq:wickno}
\end{equation}
To have a consistent nomenclature, the products of second-quantized operators $\no{\hA}\no{\hB}\no{\hC}$ are called strings, and each operator $\no{\hA}$ is called a sub-string.

\subsection{Coupled-cluster equations}
\label{subsec:cc}
To derive the CC equations, one first starts with the definition of the CC wave function which is an exponential parametrization applied to a reference determinant \cite{hubbard1957,hubbard1958}
\begin{equation}
	\ket*{ \Ps{}{(0)}} = e^{ \hT } \ket{0},
\label{eq:sche1}
\end{equation}
where $\Ps{}{(0)}$ is the CC ground-state wave function and
\begin{equation}
	\hT = \sum_{n=1}^{N}\hT_{n}
\label{eq:clust}
\end{equation}
is the cluster operator with
\begin{equation}
	\hT_{n} = \frac{1}{(n!)^2}\sum_{ijk\cdots}\sum_{abc\cdots}\at{ijk\cdots}{abc\cdots}\no{\cre{a}\ani{i}\cre{b}\ani{j}\cre{c}\ani{k}\cdots },
\label{eq:clust2}
\end{equation}
where $\at{ijk\cdots}{abc\cdots}$ are the antisymmetric cluster amplitudes and the indices $i$, $j$, $k$, $\ldots$ indicate occupied spinorbitals in the reference configuration, while $a$, $b$, $c$, $\ldots$ are unoccupied spinorbitals.
In general, $\hT_{n}$ produces $n$-fold excited $N$-electron determinants of the form $\ket*{\Phi_{ijk\cdots}^{abc\cdots}}$ (with $n \leq N$).
It is worth mentioning here that the excitation operators do commute with each other, \ie, $\comm*{\hT_{n}}{\hT_{m}} = 0$.

The electronic Schr\"odinger equation associated with the CC wave function is
\begin{equation}
	\hH \ket*{ \Ps{}{(0)}} = E^{(0)}\ket*{ \Ps{}{(0)}},
\label{eq:sche}
\end{equation}
where $E^{(0)}$ is the CC ground-state electronic energy, and $\hH$ is the electronic Hamiltonian, which can be expressed in second-quantized form as follows: 
\begin{equation}
	\hH = \sum_{pq}h_{q}^{p}\cre{p}\ani{q}+\frac{1}{4}\sum_{pqrs} \eriel{rs}{pq}  \cre{p}\cre{q}\ani{s}\ani{r},
\label{eq:hsq}
\end{equation}
where the indices $p$, $q$, $r$, and $s$ indicate arbitrary (\ie, occupied or virtual) spinorbitals. 
The matrix element $h_{q}^{p}$ is the sum of the kinetic and nuclear attraction components. 
The electronic repulsion is accounted for by the (antisymmetrized) two-electron integrals $\eriel{rs}{pq}$.
Because one is usually interested in the correlated part of the Hamiltonian, it is common practice to divide it as
\begin{equation}
	\hH = \hHN + \mel{0}{\hH}{0},
\label{eq:hsq1}
\end{equation}
where the first term 
\begin{equation}
\label{hsq1}
	\hHN 
	= \hFN + \hVN
	= \sum_{pq} \fockel{q}{p} \no{\cre{p}\ani{q}}+\frac{1}{4}\sum_{pqrs} \eriel{rs}{pq} \no{ \cre{p}\cre{q}\ani{s}\ani{r} }
\end{equation}
is the normal-ordered Hamiltonian (where $\fockel{q}{p}$ is an element of the Fock matrix) that corresponds to fluctuations (\ie, correlation) with respect to the second term that represents the reference energy $E_{0} = \bra{0}\hH\ket{0}$. It is worth emphasizing that  $\hHN$  do not commute with $\hT_{n}$, \ie, $\comm*{\hT_{n}}{\hHN} \neq 0$.

Thanks to the exponential ansatz of the CC wave function, Eq.~\eqref{eq:sche1}, and the introduction of the following CC effective Hamiltonian via a similarity transformation 
\begin{equation}
	\bH^{\text{st}} = e^{-\hT} \hHN e^{\hT},
\end{equation}
where the superindex ``$\text{st}$'' stands for ``similarity-transformed''. 
Then, one can recast the Schr\"odinger equation as
\begin{equation}
	\bH^{\text{st}} \ket{0} = \Delta E^{(0)} \ket{0}.
\label{eq:schno}
\end{equation}
where $\Delta E^{(0)}=E^{(0)}-E_{0}$ is the CC ground-state correlation energy obtained via projection 
\begin{equation}
	\mel*{0}{\bH^{\text{st}}}{0} = \Delta E^{(0)}, 
\label{eq:dele}
\end{equation}
while
\begin{equation}
	\mel*{\Ph{ijk\cdots}{abc\cdots}}{\bH^{\text{st}}}{0} = 0
\label{eq:amp}
\end{equation}
are the so-called amplitude equations, a set of non-linear equations where the unknowns are the cluster amplitudes $\at{ijk\cdots}{abc\cdots}$.

To evaluate efficiently Eq.~\eqref{eq:amp}, one usually relies on the \alert{Baker-Campbell-Hausdorff} expansion \cite{jensen2017,kumar1965} of the similarity-transformed Hamiltonian,
\begin{equation}
	\bH^{\text{st}} = \hHN +\comm{\hHN}{\hT} +\frac{1}{2!}\comm{\comm{\hHN}{\hT}}{\hT}
	+ \frac{1}{3!}\comm{\comm{\comm{\hHN}{\hT}}{\hT}}{\hT}   
	+ \frac{1}{4!}\comm{\comm{\comm{\comm{\hHN}{\hT}}{\hT}}{\hT} }{\hT},
\label{eq:hexp}
\end{equation}
a series of nested commutators between $\hHN$ and $\hT$ that naturally terminates at the four-fold commutator thanks to the two-body nature of the electronic Hamiltonian.

Then, using Wick's theorem [see Eq.~\eqref{eq:wick}], one can establish that the non-zero terms in Eq.~\eqref{eq:hexp} involve $\hHN$ as the left-most operator contracted with at most four $\hT$ operators, \ie,
\begin{equation}
	\bH^{\text{st}} = \hHN 
	+ 
	\contraction{}{\hHN}{}{\hT} 
	\hHN \hT 
	+ \frac{1}{2} 
	\contraction{}{\hHN}{}{\hT} 
	\contraction{}{\hHN}{\hT}{\hT} 
	\hHN \hT \hT  
	+ \frac{1}{3!} 
	\contraction{}{\hHN}{}{\hT} 
	\contraction{}{\hHN}{\hT}{\hT} 
	\contraction{}{\hHN}{\hT}{\hT\hT\hT}
	\hHN \hT \hT \hT 
	+ \frac{1}{4!}  
	\contraction{}{\hHN}{}{\hT} 
	\contraction{}{\hHN}{\hT}{\hT} 
	\contraction{}{\hHN}{\hT}{\hT\hT\hT} 
	\contraction{}{\hHN}{\hT}{\hT\hT\hT\hT\hT}
    \hHN \hT \hT \hT \hT,
\label{eq:hcon}
\end{equation}
where the multi-leg symbol indicates that $\hHN$ must be contracted at least once with each $\hT$. 
Hence, $\hHN$ is said to be connected as it does not include disconnected terms of the form
$\contraction{}{\hHN}{}{\hT} 
\contraction{}{\hHN}{\hT}{\hT} 
\hHN \hT \hT \hT $.

If one introduces the right-hand side of Eq.~\eqref{eq:hcon} in Eqs.~\eqref{eq:dele} and \eqref{eq:amp}, one gets
\begin{subequations}
\begin{gather}
	\mel**{ 0}{ \co{ \hHN e^{\hT} }}{ 0}  = \Delta E^{(0)},
\label{eq:dele1}
	\\
	\mel**{ \Ph{ijk\cdots}{abc\cdots}} {\co{ \hHN e^{\hT} }} {0}  = 0.
\label{eq:amp1}
\end{gather}
\end{subequations}
where the subindex $C$ stands for connected. 
Since one only deals with the connected terms, the algebraic form of Eqs.~\eqref{eq:dele1} and \eqref{eq:amp1} is simpler than Eqs.~\eqref{eq:dele} and \eqref{eq:amp}.

\subsection{Equation-of-motion coupled-cluster equations}
\label{subsec:eomcc}

Having explained how to derive the CC equations, next, we shift our attention to the EOM-CC equations. \cite{emrich1981,sekino1984,comeau1993,stanton1993} 
As a starting point, let us consider the Schrödinger equation for a target excited state $\Ps{}{(k)}$, \ie,
\begin{equation}
	\hHN  \ket*{ \Ps{}{(k)}}= \Delta E^{(k)} \ket*{ \Ps{}{(k)}},
\label{eq:schex}
\end{equation}
with $\Delta E^{(k)}= E^{(k)}-E_{0}$. 
One way to access this target state is by transforming the initial state described by Eq.~\eqref{eq:schno} via an excitation operator of the form
\begin{equation}
	\hR^{(k)} \ket*{ \Ps{}{(0)}}= \ket*{ \Ps{}{(k)}}
\label{eq:trans}
\end{equation}
If one wants to access neutral excited states, one must rely on the excitation energy (EE) operator
\begin{equation}
	\hR^{(k)} = \hI + \hR_{1}^{(k)} + \hR_{2}^{(k)} + \cdots
\label{eq:excop}
\end{equation}
with
\begin{subequations}
\begin{gather}
	\hR_{1}^{(k)} = \sum_{i}\sum_{a}\ar{i}{a}\no{\cre{a}\ani{i}},
	\\
	\hR_{2}^{(k)} = \frac{1}{4}\sum_{ij}\sum_{ab}\ar{ij}{ab}\no{\cre{a}\ani{i}\cre{b}\ani{j}},
	\\
	\vdots \notag
\label{eq:excop1}
\end{gather}
\end{subequations}
where $\hR_{n}^{(k)}$ is an excitation operator of degree $n$ for the state $k$ associated with the EOM-CC amplitudes $\ar{ab\cdots}{ij\cdots}$.
(Here, for the sake of simplicity, we have omitted the $k$-dependence of the amplitudes.)

After several transformations, the Schr\"odinger equation for the target excited state [see Eq.~\eqref{eq:schex}] can be recast as
\begin{equation}
	\comm{ \bH^{\text{st}}}{  \hR^{(k)}} \ket{ 0}= \Omega_{k} \hR^{(k)} \ket{0}.
\label{eq:eomcccomm}
\end{equation}
where $\Omega_{k} =  E^{(k)} - E^{(0)}$ is the excitation energy associated with the $k$th excited state.
Defining the normal-ordered similarity-transformed Hamiltonian, $\bHN^{\text{st}} = \alert{\bH^{\text{st}}} - E^{(0)}$,
the commutator in Eq.~\eqref{eq:eomcccomm} can be further simplified as
\begin{equation}
	\co{ \bHN^{\text{st}}  \hR^{(k)}} \ket{0}= \Omega_{k} \hR^{(k)} \ket{ 0}.
\label{eq:eomcc}
\end{equation}
The excitation energies are then directly computed by projecting Eq.~\eqref{eq:eomcc} in a determinant basis truncated at a given excitation degree, which is equivalent to truncate $\hR^{(k)}$.

Introducing the simplified notation, $\ket{\Ph{i}{a}} \equiv \ket{S}$, $\ket{\Ph{ij}{ab}} \equiv \ket{D}$, $\ldots$, the EOM-CC linear eigenvalue problem has the following form:
\begin{equation}
\bEOM\bold{r}_{k}=\Omega_{k}\bold{r}_{k}
\qq{$\Rightarrow$}
\begin{pmatrix}
 \bra{S}\bHN^{\text{st}}\ket{S} & \bra{ S}\bHN^{\text{st}}\ket{D}& \cdots \\
 \bra{D}\bHN^{\text{st}}\ket{S} & \bra{ D}\bHN^{\text{st}}\ket{D}& \cdots  \\
 \vdots & \vdots   & \ddots                               
\end{pmatrix}
\begin{pmatrix}
\ar{i}{a}& \\
\ar{ij}{ab} &\\
\vdots  &  
\end{pmatrix}
=
\Omega_{k}
\begin{pmatrix}
\ar{i}{a}& \\
\ar{ij}{ab} &\\
\vdots  &  
\end{pmatrix}.
\label{eq:eomccmf}
\end{equation}

\alert{Here, $\bEOM$ represents the EOM-CC matrix, while $\bold{r}_{k}$ and $\Omega_{k}$ are the corresponding eigenvector and eigenvalue, respectively. The non-Hermitian nature of this eigenvalue problem necessitates the computation of each block of the matrix. Furthermore, this non-Hermitian characteristic indicates that $\bEOM$ is also associated with left eigenfunctions $ \bra*{ \Ps{}{(0)}} \hL_{1}^{(k)}= \bra*{ \Ps{}{(k)}} $ stemming from the following Schrödinger equation $ \bra*{ \Ps{}{(k)}}\hHN = \bra*{ \Ps{}{(k)}}\Delta E^{(k)}$, where $\hL^{(k)} = \hI + \hL_{1}^{(k)} + \hL_{2}^{(k)} + \cdots$ is a de-excitation operator, such that $\hL_{1}^{(1)} = \sum_{i}\sum_{a}\al{i}{a}\no{\cre{i}\ani{a}}$, for example.  This results in the following left-hand side eigenvalue equation
\begin{equation}
	\bold{l}_{k}\bEOM=\bold{l}_{k}\Omega_{k}^{*}
	\label{eq:eomccmfLeft}
\end{equation}
The two sets of eigenfunctions can be normalized to satisfy $\bra*{ \Ps{}{(0)}} \hL_{1}^{(l)} \hR^{(k)} \ket*{ \Ps{}{(0)}}=\delta_{lk}$.}

Following the same procedure, one can also obtain the EOM-CC equations for the ionized and electron-attached states, just by changing the definition of the operator $\hR^{(k)}$ in Eq.~\eqref{eq:excop1}. 
For example, removing one or two particles, one gets IP-EOM-CC and DIP-EOM-CC, \cite{sattelmeyer2003,demel2008,musial2011,kus2011,shen2013} respectively, while removing one or two holes yields EA-EOM-CC and DEA-EOM-CC, \cite{kus2011,shen2013} respectively. 

\subsection{Many-body Hamiltonian}
\label{subsec:mbodh}

When one computes the expectation value associated with a given block of the EOM-CC matrix $\bEOM$ [see Eq.~\eqref{eq:eomccmf}], redundant terms are generated. 
One way to avoid redundancies is to rely on the many-body representation of the EOM-CC effective Hamiltonian \cite{kucharski1991,kucharski1992,musial2002,musial2002a} 
\begin{equation}
	\bHN^{\text{mb}} = \sum_{pq}\mbt{q}{p}\no{\cre{p}\ani{q}}+ \sum_{pqrs}\mbt{rs}{pq}\no{\cre{p}\cre{q}\ani{s}\ani{r}}
	+ \mbt{stu}{pqr}\no{\cre{p}\cre{q}\cre{r}\ani{u}\ani{t}\ani{s}} + \cdots,
\label{eq:effham}
\end{equation}
where  $\mbt{q}{p}$, $\mbt{rs}{pq}$, $\mbt{stu}{pqr}$ are one-, two-, and three-body terms respectively. The superindex ``$\text{mb}$'' stands for ``many-body''. 
Note that $\bHN^{\text{mb}}$ and $\bHN^{\text{st}}$ are exactly the same quantity; the labels ``\text{mb}'' and ``\text{st}'' are here to indicate the type of expansion.

The many-body terms $\{\mbt{stu\cdots}{pqr\cdots}\}$ are usually represented through diagrammatic techniques. \cite{shavitt2009} 
In the present work though, we are interested in describing them within the second quantization formalism. 
To achieve this, we take as an example the block $\bra{D}\bHN^{\text{mb}}\ket{S}$ and the many-body Hamiltonian defined in Eq.~\eqref{eq:effham} to obtain
\begin{multline}
	\bra{\Ph{ij}{ab}}\bHN^{\text{mb}}\ket{\Ph{k}{c}}
	= \sum_{pq}\mbt{q}{p}\bra{0} \no{\cre{i}\cre{j}\ani{b}\ani{a}}\no{\cre{p}\ani{q}}\no{\cre{c}\ani{k}}\ket{0}
	+ \sum_{pqrs}\mbt{rs}{pq}\bra{0}\no{\cre{i}\cre{j}\ani{b}\ani{a}}\no{\cre{p}\cre{q}\ani{s}\ani{r}}\no{\cre{c}\ani{k}}\ket{0}
\\
	 + \sum_{pqrstu}\mbt{stu}{pqr}\bra{0}\no{\cre{i}\cre{j}\ani{b}\ani{a}}\no{\cre{p}\cre{q}\cre{r}\ani{u}\ani{t}\ani{s}}\no{\cre{c}\ani{k}}\ket{0}.
\label{eq:string2}
\end{multline}
Equation \eqref{eq:string2} effectively terminates at the three-body level since higher-order terms produce partial contractions which are zero with respect to the Fermi vacuum.

By applying Wick's theorem to Eq.~\eqref{eq:string2} and taking advantage of the antisymmetric permutation of the many-body terms, one gets
\begin{multline}
   \mel**{\Ph{ij}{ab}}{\bHN^{\text{mb}}}{\Ph{k}{c}} = \delta_{jk}\delta_{bc}\mbt{i}{a}
	- \delta_{ik}\delta_{bc}\mbt{j}{a}
	- \delta_{jk}\delta_{ac}\mbt{i}{b}
	+ \delta_{jk}\delta_{ac}\mbt{j}{b}
	- \delta_{bc}\mbt{ij}{ak}
	- \delta_{jk}\mbt{ci}{ab}
	+ \delta_{ik}\mbt{cj}{ab}
	+ \delta_{ac}\mbt{ij}{bk}
	 + \mbt{ijc}{abk}.
\label{eq:str3}
\end{multline}
Since $\bHN^{\text{mb}} = \bHN^{\text{st}}$, the block $\bra{D}\bHN^{\text{mb}}\ket{S}$ can be naturally written in terms of $\hT$, $\hFN$, and $\hVN$. 
To illustrate this, we consider the general many-body term
\begin{equation}
	\mbt{ij\cdots cd\cdots}{ab\cdots kl\cdots}=\bra{\Ph{ij\cdots}{ab\cdots}}\bHN^{\text{st}}\ket{\Ph{kl\cdots}{cd\cdots}}_{iC},
\label{eq:gmbt}
\end{equation} 
where the subscript $iC$ represents ``internal contractions'', \ie, contractions occurring exclusively between $\bra{\Ph{ij\cdots}{ab\cdots}}$ and $\bHN$, as well as $\bHN$ and $\ket{\Ph{kl\cdots}{cd\cdots}}$. 
Notably, contractions between $\bra{\Ph{ij\cdots}{ab\cdots}}$ and $\ket{\Ph{kl\cdots}{cd\cdots}}$ are not allowed.

Taking into account Eq.~\eqref{eq:gmbt}  and the non-zero matrix elements of $\mbt{ij}{ak}$, one gets
\begin{equation}
	\mbt{ij}{ak}=
	\mel**{\Ph{ij}{a}} {\qty(\hVN+\hT_{1}\hVN +\hFN\hT_{2}+
	\hVN\hT_{2}+ \frac{1}{2!}\hVN\hT_{1}^2+\hVN\hT_{1}\hT_{2}+ \frac{1}{3!}\hVN\hT_{1}^3)_{C}} {\Ph{k}{}}_{iC}.
\label{eq:intermed}
\end{equation}
Applying the same procedure for $\mbt{ci}{ab}$ and $\mbt{ijc}{abk}$, we obtain
\begin{gather}
	\mbt{ic}{ab} 
	= \mel**{\Ph{i}{ab}} {\qty( \hVN+\hT_{1}\hVN +\hFN\hT_{2}	
	+ \hVN\hT_{2}+ \frac{1}{2!}\hVN\hT_{1}^2+\hVN\hT_{1}\hT_{2}+ \frac{1}{3!}\hVN\hT_{1}^3)_{C}} {\Ph{}{c}}_{iC},
\label{eq:intermed2}
	\\
	\mbt{ijc}{abk} = \mel**{\Ph{ij}{ab}}{\left(\hVN \hT_{2}\right)_{C}}{\Ph{k}{c}}_{iC}.
\label{eq:intermed3}
\end{gather}
The one-body terms in Eq.~\eqref{eq:str3} are defined as follows
\begin{equation}
	\mbt{i}{a}
	= \mel**{\Ph{i}{a}}
	{\qty[ \hHN \qty( 1+\hT_{1} +\hT_{2}+\frac{\hT_{1}^2}{2}
	+\hT_{1}\hT_{2}+\frac{\hT_{1}^3}{3!} ) ]_{C}}{0},
\label{eq:ccsdhamiltonian}
\end{equation}
which is identical to the CCS amplitude equations. 
Consequently, all the one-body terms are zero in Eq.~\eqref{eq:str3}. 
This is also true for the many-body terms $\mbt{ij}{ab}$ for CCSD, $\mbt{ijk}{abc}$ for CCSDT, and so on. 

Setting the one-body terms equal to zero in Eq.~\eqref{eq:ccsdhamiltonian}, we have
\begin{equation}
\bra{\Ph{ij}{ab}}\bHN^{\text{mb}}\ket{\Ph{k}{c}}=
	- \delta_{bc}\mbt{ij}{ak}
	- \delta_{jk}\mbt{ci}{ab}
	+ \delta_{ik}\mbt{cj}{ab}
	+ \delta_{ac}\mbt{ij}{bk}
	+ \mbt{ijc}{abk}.
\label{eq:str4}
\end{equation}
It is possible to obtain the rest of the many-body terms in Eq.~\eqref{eq:str4} by substituting the indices of $\mbt{ij}{ab}$, $\mbt{ci}{ab}$, and $\mbt{ijc}{abk}$. 
These replacements are restricted to indices of the same category: hole-creation ($i,j$), particle-annihilation ($a,b$), hole-annihilation ($k$), and particle-creation ($c$).

To obtain the block $\bra{\Ph{ij}{ab}}\bHN^{\text{mb}}\ket{\Ph{k}{c}}$ in a compact and non-redundant form, we rely on Eqs.~\eqref{eq:intermed}, \eqref{eq:intermed2}, and \eqref{eq:intermed3} to define the many-body terms and Eq.~\eqref{eq:str4} to define the blocks in terms of these many-body terms. 
Finally, to generate the entire eigenvalue equations for EE-EOM-CCSD, one must repeat the same procedure for the blocks $\bra{\Ph{i}{a}}\bHN^{\text{st}}\ket{\Ph{k}{c}}$, $\bra{\Ph{i}{a}}\bHN^{\text{st}}\ket{\Ph{kl}{cd}}$, $\bra{\Ph{ij}{ab}}\bHN^{\text{st}}\ket{\Ph{kl}{cd}}$, and then remove the redundant many-body terms.

\section{EOM-CC equation generator}
\label{sec:eomcceqgen}

In this section, we describe the algorithm we have written with \mat to automatically derive the working equations of EOM-CC. 
Although the program is not optimized for efficiency, our goal is to obtain the working equations in terms of the many-body terms by using straightforward input quantities. 
First of all, we adapt the index notations in order to have more suitable and general notations to obtain the EOM-CC equations up to fourth order. 
For the operators belonging to the bra and the ket, instead of using $\cre{a}$, $\cre{b}$, $\ldots$, $\ani{i}$, $\ani{j}$, $\ldots$, we use $\crn{p}{1}$, $\crn{p}{2}$, $\ldots$, $\ann{h}{1}$, $\ann{h}{2}$, $\ldots$. 
For the particles/holes that play the role of dummy indices in the cluster operator $\hT$, instead of using $\cre{e}$, $\cre{f}$, $\ldots$, $\ani{m}$, $\ani{n}$, $\ldots$, we use $\crn{o}{1}$, $\crn{o}{2}$, $\ldots$, $\ann{v}{1}$, $\ann{v}{2}$, $\ldots$, where the notation of $\ind{o}{}$ and $\ind{v}{}$ refers to occupied and virtual, respectively.
Finally, for the arbitrary indexes that could be either particle or hole [like the Hamiltonian in Eq.~\eqref{eq:hsq1}], we switch from $\cre{p}$, $\cre{q}$, $\ani{r}$, $\ani{s}$ to $\crn{q}{1}$, $\crn{q}{2}$, $\ann{q}{3}$, $\ann{q}{4}$. 

Thanks to this change in notation, we can now describe the four-step algorithm that we use to derive the EOM-CC equations. 
Each step of the algorithm is written as a \mat module. 
These modules are gathered within a main module called \red{\texttt{eomccgen}}. 
Variables tailored-made for \mat are summarized in Table \ref{table:compnotation} alongside their description and corresponding mathematical expression.

\begin{table*}
\caption{List of variables used in the EOM-CC equation generator program.}
\label{table:compnotation} 
\begin{ruledtabular}
\begin{tabular}{lll} 
 Description  & Mathematical symbol & \mat notation  \\ 
 \hline 
	Elements of the Fock matrix				&	$\fockel{q}{p}$		&	$\texttt{F}[\![\texttt{p,q} ]\!]$				\\ 
	Two-electron repulsion integrals		&	$\eriel{rs}{pq}$	&	$\texttt{ERI}[\![\texttt{p,q,r,s} ]\!]$			\\
	Amplitudes of single excitations		&	$\at{i}{a}$			&	$\texttt{t1}[\![\texttt{i,a} ]\!]$				\\ 
	Amplitudes of double excitations		&	$\at{ij}{ab}$		&	$\texttt{t2}[\![\texttt{i,j,a,b} ]\!]$			\\
	Amplitudes of triple excitations		&	$\at{ijk}{abc}$		&	$\texttt{t3}[\![\texttt{i,j,k,a,b,c} ]\!]$		\\
	Amplitudes of quadruple excitations		&	$\at{ijkl}{abcd}$	&	$\texttt{t4}[\![\texttt{i,j,k,l,a,b,c,d} ]\!]$	\\
	One-body terms							&	$\mbt{q}{p}$		&	$\chi1[\![\texttt{p,q} ]\!]$					\\
	Two-body terms							&	$\mbt{rs}{pq}$		&	$\chi2[\![\texttt{p,q,r,s} ]\!]$				\\
	Three-body terms						&	$\mbt{stu}{pqr}$	&	$\chi3[\![\texttt{p,q,r,s,t,u} ]\!]$			\\
	Four-body terms							&	$\mbt{tuvw}{pqrs}$	&	$\chi4[\![\texttt{p,q,r,s,t,u,v,w} ]\!]$		\\
	EOMCC right-hand amplitudes of single excitations		&	$\ar{i}{a}$			&	$\texttt{r1}[\![\texttt{i,a} ]\!]$				\\ 
	EOMCC right-hand amplitudes of double excitations		&	$\ar{ij}{ab}$		&	$\texttt{r2}[\![\texttt{i,j,a,b} ]\!]$			\\
	EOMCC right-hand amplitudes of triple excitations		&	$\ar{ijk}{abc}$		&	$\texttt{r3}[\![\texttt{i,j,k,a,b,c} ]\!]$		\\
	EOMCC right-hand amplitudes of quadruple excitations		&	$\ar{ijkl}{abcd}$	&	$\texttt{r4}[\![\texttt{i,j,k,l,a,b,c,d} ]\!]$	\\
		EOMCC left-hand amplitudes of single excitations		&	$\al{i}{a}$			&	$\texttt{l1}[\![\texttt{i,a} ]\!]$				\\ 
	EOMCC left-hand amplitudes of double excitations		&	$\al{ij}{ab}$		&	$\texttt{l2}[\![\texttt{i,j,a,b} ]\!]$			\\
	EOMCC left-hand amplitudes of triple excitations		&	$\al{ijk}{abc}$		&	$\texttt{l3}[\![\texttt{i,j,k,a,b,c} ]\!]$		\\
	EOMCC left-hand amplitudes of quadruple excitations		&	$\al{ijkl}{abcd}$	&	$\texttt{l4}[\![\texttt{i,j,k,l,a,b,c,d} ]\!]$	\\
\end{tabular}
\end{ruledtabular}
\end{table*}

\begin{figure*}
	\includegraphics[width=\textwidth]{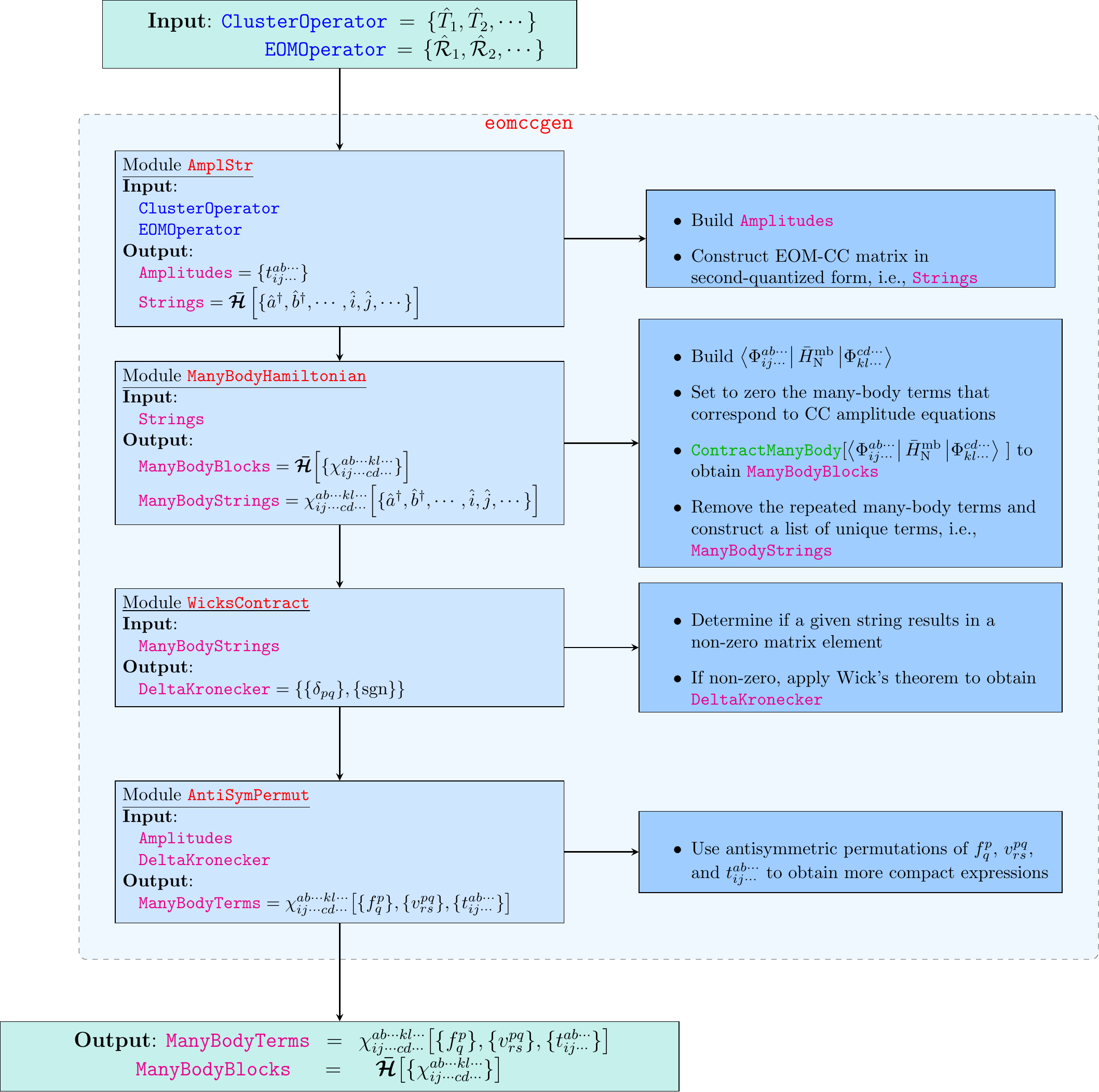}
	\caption{Flowchart of the entire set of modules constituting the EOM-CC equation generator.}
	\label{Fig:eomccalg} 
\end{figure*}

The input quantities for \red{\texttt{eomccgen}} are the list of (neutral) excitation operators, named \blue{\texttt{ClusterOperator}} (that determine the similarity-transform Hamiltonian) and the list of EOM operators, named \blue{\texttt{EOMOperator}}.  
For example, the input to obtain the EE-EOM-CCSDT equations is
\begin{equation*}
\begin{split}
	\blue{\texttt{ClusterOperator}} & = \texttt{ \{\{\purple{"1h1p"}\},\{\purple{"2h2p"}\},\{\purple{"3h3p"}\}\}}
	\\
	\blue{\texttt{EOMOperator}} & = \texttt{ \{\{\purple{"1h1p"}\},\{\purple{"2h2p"}\},\{\purple{"3h3p"}\}\}}
\end{split}
\end{equation*}
If one wants to get IP-EOM-CCSD, the input is 
\begin{equation*}
\begin{split}
	\blue{\texttt{ClusterOperator}} & = \texttt{ \{\{\purple{"1h1p"}\},\{\purple{"2h2p"}\}\}}
	\\
	\blue{\texttt{EOMOperator}} & = \texttt{ \{\{\purple{"1h0p"}\},\{\purple{"2h1p"}\}\}\}}
\end{split}
\end{equation*}
To obtain DEA-EOM-CCSD, the second line has to be changed as
\begin{equation*}
	\blue{\texttt{EOMOperator}} = \texttt{ \{\{\purple{"0h2p"}\},\{\purple{"1h3p"}\}\}\}}
\end{equation*}
It is also possible to have different numbers of operators in \blue{\texttt{EOMOperator}} and \blue{\texttt{ClusterOperator}}. 
For example, the EA-EOM-CCSD(3h2p) equations can be produced as
\begin{equation*}
	\blue{\texttt{EOMOperator}} = \texttt{ \{\{\purple{"0h1p"}\},\{\purple{"1h2p"}\},\{\purple{"3h2p"}\}\}}
\end{equation*}
In practice, it is possible to carry out all possible combinations of excitation operators up to 4h4p, and it is also possible to calculate the CC amplitude equations. 
For example, to generate the CCSD amplitude equations, the following input must be entered
\begin{equation*}
\begin{split}
	\blue{\texttt{ClusterOperator}} & = \texttt{ \{\{\purple{"1h1p"}\},\{\purple{"2h2p"}\}\}}
	\\
	\blue{\texttt{EOMOperator}} & = \texttt{ \{\{\purple{"0h0p"}\}\}}
\end{split}
\end{equation*}
We shall now discuss in detail the algorithm summarized in Fig.~\ref{Fig:eomccalg}.

\subsection{Step 1}
\label{subsec:stp1}
The first step is carried out through the module \red{\texttt{AmplStr}}. 
The inputs of this module are the \blue{\texttt{ClusterOperator}} and \blue{\texttt{EOMOperator}}:
\begin{equation*}
	\{\brown{\texttt{Amplitudes}},\brown{\texttt{Strings}}\}=
	\red{\texttt{AmplStr}}\qty[\blue{\texttt{ClusterOperator}},\blue{\texttt{EOMOperator}} ]
\end{equation*}
It returns the list of CC amplitudes (stemming from the truncated Taylor expansion of $e^{\hT}$) in \brown{\texttt{Amplitudes}}.
For example, in the case of EE-EOM-CCSD, we have
\begin{equation*}
    \brown{\texttt{Amplitudes}}=\{1,\at{\ind{o}{1}}{\ind{v}{1}},\at{\ind{o}{1}}{\ind{v}{1}}\at{\ind{o}{2}}{\ind{v}{2}},\at{\ind{o}{1}}{\ind{v}{1}}\at{\ind{o}{2}}{\ind{v}{2}}\at{\ind{o}{3}}{\ind{v}{3}},\at{\ind{o}{1}\ind{o}{2}}{\ind{v}{1}\ind{v}{2}},\at{\ind{o}{1}\ind{o}{2}}{\ind{v}{1}\ind{v}{2}}\at{\ind{o}{3}}{\ind{v}{3}}\}
\end{equation*}
while, in $\mat$, it takes the following form:
\begin{lstlisting}[extendedchars=true,language=Mathematica,mathescape=true]
	Amplitudes={{1},{t1$[\![$o1,v1$]\!]$},{t1$[\![$o1,v1$]\!]$t1$[\![$o2,v2$]\!]$},{t1$[\![$o1,v1$]\!]$t1$[\![$o2,v2$]\!]$t1$[\![$o3,v3$]\!]$},{t2$[\![$o1,o2,v1,v2$]\!]$},{t1$[\![$o1,v1$]\!]$t2$[\![$o2,o3,v2,v3$]\!]$}}
\end{lstlisting}
The set of excitation operators gathered in \blue{\texttt{EOMOperator}} determines the basis in which $\bEOM$ [see Eq.~\eqref{eq:eomccmf}] is diagonalized. 
With this basis together with the similarity-transformed Hamiltonian, the EOM-CC matrix is written in second-quantized form, $\bEOM\qty[\{\crn{p}{1},\crn{p}{2},\cdots,\ann{h}{1},\ann{h}{2},\cdots\} ]$, and stored in the variable \brown{\texttt{Strings}}, which is made of two substrings, i.e., $\brown{\texttt{Strings}}=\{\brown{\texttt{StringF}},\brown{\texttt{StringV}} \}$, where $\brown{\texttt{StringF}}$ (F stands for $\hFN$) and $\brown{\texttt{StringV}}$ (V stands for $\hVN$).  
For example, the element $\brown{\texttt{StringV}}[\![\texttt{2,1,6} ]\!]$ corresponds to
\begin{equation*}
    \bra{D}\hVN\hT_{1}\hT_{2}\ket{S}=
    \\
    \texttt{\{}
    \texttt{\{}\texttt{h1}^{\dag}\texttt{,}\texttt{h2}^{\dag} \texttt{,p2,p1\}} \texttt{,}
    \texttt{\{}\texttt{q1}^{\dag}\texttt{,}\texttt{q2}^{\dag}\texttt{,q4,q3\}} \texttt{,}  
    \\
    \texttt{\{}\texttt{\{}\texttt{v1}^{\dag}\texttt{,}\texttt{o1\}\texttt{,}\{}  \texttt{v2}^{\dag}\texttt{,}\texttt{o2} \texttt{,}\texttt{v3}^{\dag}\texttt{,}\texttt{o3\}\}}\texttt{,}
    \texttt{\{}\texttt{p3}^{\dag}\texttt{,}\texttt{h3\}}
    \texttt{\}}
\end{equation*}
The values \texttt{2} and \texttt{1} in $\brown{\texttt{StringV}}[\![\texttt{2,1,6} ]\!]$ correspond to the indices of the row and column of the EOM-CC matrix, respectively, while \texttt{6} represents the sixth amplitude, i.e., $\brown{\texttt{Amplitudes}}[\![\texttt{6}]\!]$.
For the sake of simplicity, the summation and the Fermi vacuum of \brown{\texttt{StringF}} and \brown{\texttt{StringV}} are implicit.

\subsection{Step 2}
\label{subsec:stp2}

In Fig.~\ref{Fig:eomccalg}, the module that corresponds to the second step is named \red{\texttt{ManyBodyHamiltonian}} and requires the input quantity \brown{\texttt{Strings}} generated in \red{\texttt{AmplStr}} (see Subsec.~\ref{subsec:stp1}):
\begin{equation*}
	\{\brown{\texttt{ManyBodyBlocks}},\brown{\texttt{ManyBodyStrings}}\}=
	\red{\texttt{ManyBodyHamiltonian}}\qty[\brown{\texttt{Strings}} ]
\end{equation*}
where \brown{\texttt{ManyBodyBlocks}} corresponds to the EOM-CC matrix defined in Eq.~\eqref{eq:eomccmf} written in terms of the many-body terms, i.e., $\bEOM\qty[ \{ \mbt{\ind{h}{1}\ind{h}{2}\cdots \ind{p}{3}\ind{p}{4}\cdots}{\ind{p}{1}\ind{p}{2}\cdots \ind{h}{3}\ind{h}{4}\cdots} \} ]$, and \brown{\texttt{ManyBodyStrings}} is a list that consists of all unique many-body terms in second-quantized form, i.e., $\mbt{\ind{h}{1}\ind{h}{2}\cdots \ind{p}{3}\ind{p}{4}\cdots}{\ind{p}{1}\ind{p}{2}\cdots \ind{h}{3}\ind{h}{4}\cdots} \qty[\{\crn{p}{1},\crn{p}{2},\cdots,\ann{h}{1},\ann{h}{2},\cdots\} ]$. 

More precisely, the algorithm inside \red{\texttt{ManyBodyHamiltonian}} that generates \brown{\texttt{ManyBodyBlocks}} and \brown{\texttt{ManyBodyStrings}} performs the following steps:

\begin{enumerate}

	\item Extract the EOM basis from \brown{\texttt{Strings}} and evaluate the many-body Hamiltonian represented in Eq.~\eqref{eq:effham} to form matrix elements, such as $\bra{\Ph{\ind{h}{1}\ind{h}{2}\cdots}{\ind{p}{1}\ind{p}{2}\cdots}}\bHN^{\text{mb}}\ket{\Ph{\ind{h}{3}\ind{h}{4}\cdots}{\ind{p}{3}\ind{p}{4}\cdots}}$.

	\item Set to zero the many-body terms that correspond to the CC amplitude equations, such as the ones defined in Eq.~\eqref{eq:ccsdhamiltonian}.

	\item Contract blocks made of second-quantized strings of the form $\bra{\Ph{\ind{h}{1}\ind{h}{2}\cdots}{\ind{p}{1}\ind{p}  {2}\cdots}}\bHN^{\text{mb}}\ket{\Ph{\ind{h}{3}\ind{h}{4}\cdots}{\ind{p}{3}\ind{p} {4}\cdots}}$ with the help of the \green{\texttt{ContractManyBody}} function. 
	For example, the contraction of the block $\bra{\Ph{\ind{h}{1}}{\ind{p}{1}}}\bHN^{\text{mb}}\ket{\Ph{\ind{h}{3}}{\ind{p}{3}}}$ generates
    \begin{equation*}
        (-\delta_{p_1 p_3}\mbt{\ind{h}{1}}{\ind{h}{3}}, 
        \delta_{h_1 h_3}\mbt{\ind{p}{3}}{\ind{p}{1}},
        - \mbt{\ind{h}{1}\ind{p}{3}}{\ind{p}{1}\ind{h}{3}} )=
        \green{\texttt{ContractManyBody}}[\bra{\Ph{\ind{h}{1}}{\ind{p}{1}}}\bHN^{\text{mb}}\ket{\Ph{\ind{h}{3}}{\ind{p}{3}}}]
    \end{equation*}
	Applying  \green{\texttt{ContractManyBody}} to each block of $\bEOM$ allows us to obtain the EOM-CC matrix in terms of many-body terms, i.e., $\bEOM\qty[\{ \mbt{\ind{h}{1}\ind{h}{2}\cdots \ind{p}{3}\ind{p}{4}\cdots}{\ind{p}{1}\ind{p}{2}\cdots \ind{h}{3}\ind{h}{4}\cdots} \}]$.
	These are stored in \brown{\texttt{ManyBodyBlocks}}. 
	In $\mat$, it takes the following form:
\begin{lstlisting}[extendedchars=true,language=Mathematica,mathescape=true]
	ManyBodyBlocks={-$\delta_{\texttt{p1,p3}}$$\chi$1$[\![$h3,h1$]\!]$+$\delta_{\texttt{h1,h3}}$$\chi$1$[\![$p1,p3$]\!]$-$\chi$2$[\![$h3,p1,h1,p3$]\!]$}
\end{lstlisting}

	\item For each block of $\brown{\texttt{ManyBodyBlocks}}$, we remove any redundant many-body term to create a list of unique terms, $\mbt{\ind{h}{1}\ind{h}{2}\cdots \ind{p}{3}\ind{p}{4}\cdots}{\ind{p}{1}\ind{p}{2}\cdots \ind{h}{3}\ind{h}{4}\cdots} \qty[\{\crn{p}{1},\crn{p}{2},\cdots,\ann{h}{1},\ann{h}{2},\cdots\} ]$, stored in \brown{\texttt{ManyBodyStrings}}.
	This is done by counting the number of hole-annihilation, hole-creation, particle-annihilation, and particle-creation operators in each term. 
	For example, in the case of IP-EOM-CCSD, we have
    \begin{equation*}
        \brown{\texttt{ManyBodyStrings}} =\{\mbt{\ind{h}{1}}{\ind{h}{3}},\mbt{\ind{p}{3}}{\ind{h}{4}}, \mbt{\ind{p}{3}}{\ind{p}{1}},
        \mbt{\ind{h}{1}\ind{h}{2}}{\ind{h}{3}\ind{p}{1}}, 
        \mbt{\ind{h}{2}\ind{p}{3}}{\ind{h}{4}\ind{p}{1}},
        \mbt{\ind{h}{1}\ind{h}{2}\ind{p}{3}}{\ind{h}{3}\ind{h}{4}\ind{p}{1}}
        \}
    \end{equation*}
	Next, we express each many-body term as in Eq.~\eqref{eq:gmbt} to obtain their second-quantized form. 
	For example, 
\begin{equation}
	\mbt{\ind{h}{1}\ind{h}{2}\ind{p}{3}}{\ind{h}{3}\ind{h}{4}\ind{p}{1}}=\bra{\Ph{\ind{h}{1}\ind{h}{2}}{\ind{p}{1}}}\bHN^{\text{st}}\ket{\Ph{\ind{h}{3}\ind{h}{4}}{\ind{p}{3}}}_{iC}
\end{equation} 
	while, in $\mat$, this three-body term corresponds to the sixth position of $\brown{\texttt{ManyBodyStrings}}$:
\begin{lstlisting}[extendedchars=true,language=Mathematica,mathescape=true]
	ManyBodyBlocks$[\![$6$]\!]$={{{SuperDagger[h1],SuperDagger[h2],p1},{SuperDagger[q1],SuperDagger[q2],q4,q3},{},{SuperDagger[p3],h4,h3}},{{SuperDagger[h1],SuperDagger[h2],p1},{SuperDagger[q1],SuperDagger[q2],q4,q3},{{SuperDagger[v1],o1}},{SuperDagger[p3],h4,h3}},{{SuperDagger[h1],SuperDagger[h2],p1},{SuperDagger[q1],SuperDagger[q2],q4,q3},{{SuperDagger[v1],o1},{SuperDagger[v2],o2}},{SuperDagger[p3],h4,h3}},{{SuperDagger[h1],SuperDagger[h2],p1},{SuperDagger[q1],SuperDagger[q2],q4,q3},{{SuperDagger[v1],o1},{SuperDagger[v2],o2},{SuperDagger[v3],o3}},{SuperDagger[p3],h4,h3}},{{SuperDagger[h1],SuperDagger[h2],p1},{SuperDagger[q1],SuperDagger[q2],q4,q3},{{SuperDagger[v1],o1,SuperDagger[v2],o2}},{SuperDagger[p3],h4,h3}},{{SuperDagger[h1],SuperDagger[h2],p1},{SuperDagger[q1],SuperDagger[q2],q4,q3},{{SuperDagger[v1],o1},{SuperDagger[v2],o2,SuperDagger[v3],o3}},{SuperDagger[p3],h4,h3}}}\end{lstlisting}

\end{enumerate}

\subsection{Step 3}
\label{subsec:stp3}
The third step is carried out by the module named \red{\texttt{WicksContract}}, which performs Wick's contractions [see Eq.~\eqref{eq:wickno}] of the second-quantized strings in \brown{\texttt{ManyBodyStrings}} obtained in the module \red{\texttt{ManyBodyHamiltonian}}, as follows:
\begin{equation*}
	\brown{\texttt{DeltaKronecker}}=
	\\
	\red{\texttt{WicksContract}}[\brown{\texttt{ManyBodyStrings}} ]
\end{equation*}
The output \brown{\texttt{DeltaKronecker}} contains two quantities, the Kronecker deltas and the signs of each contraction.

\begin{figure*}
	\includegraphics[width=0.6\textwidth]{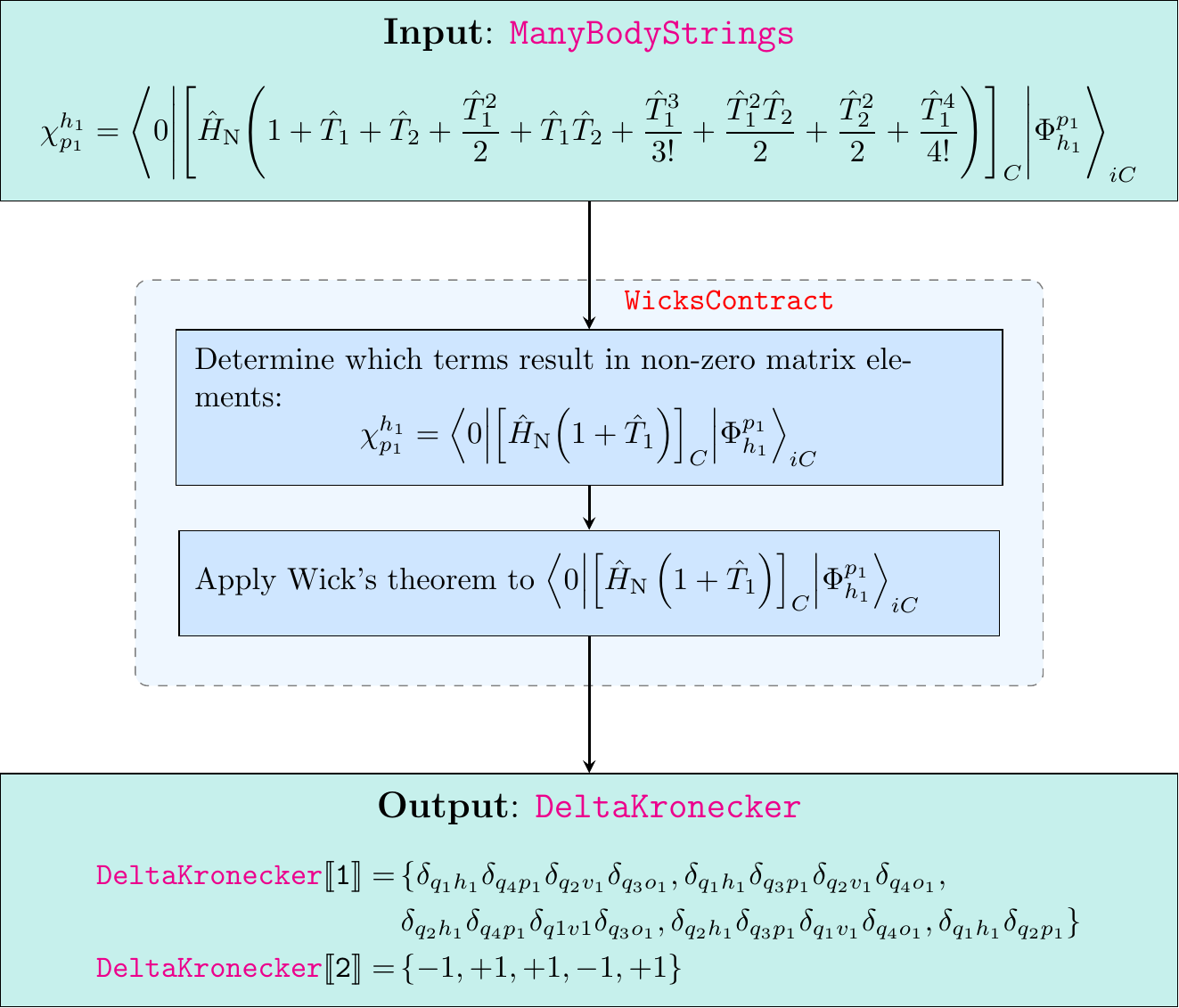}
	\caption{Flowchart of the module \red{\texttt{WicksContract}}. 
	The module takes as input $\mbt{\ind{p}{1}}{\ind{h}{1}}=\mel**{0} {\bHN} {\Ph{\ind{h}{1}}{\ind{p}{1}}}_{iC}$ stored in the variable \brown{\texttt{ManyBodyStrings}} produced from the module \red{\texttt{ManyBodyHamiltonian}} (see Fig.~\ref{Fig:eomccalg}).}
	\label{Fig:wickscontract} 
\end{figure*}

Figure \ref{Fig:wickscontract} illustrates the structure of the module \red{\texttt{WicksContract}}. 
Its input is the one-body term $\mbt{\ind{p}{1}}{\ind{h}{1}}$ which is stored in the variable $\brown{\texttt{ManyBodyStrings}}[\![\texttt{3}]\!]$ produced by the module \red{\texttt{ManyBodyHamiltonian}} discussed in Subsec.~\ref{subsec:stp2}. 
The first step consists of replacing the similarity-transformed Hamiltonian with the terms yielding non-zero matrix elements and then applying Wick's theorem to the remaining terms. 
Finally, the Kronecker deltas and signs obtained from Wick's theorem are stored in \brown{\texttt{DeltaKronecker}}. 

\subsection{Step 4}
\label{subsec:stp4}

The module \red{\texttt{AntiSymPermut}} requires three inputs: \brown{\texttt{DeltaKronecker}}, \brown{\texttt{ManyBodyStrings}}, and \brown{\texttt{Amplitudes}}.
\begin{equation*}
	\brown{\texttt{ManyBodyTerms}}=
	\red{\texttt{AntiSymPermut}}[\brown{\texttt{DeltaKronecker}},\brown{\texttt{ManyBodyStrings}},\brown{\texttt{Amplitudes}} ]
\end{equation*}
This module applies the Kronecker deltas obtained in \red{\texttt{WicksContract}} (see Subsec.~\ref{subsec:stp3}) to the \brown{\texttt{Amplitudes}} (generated in \red{\texttt{AmplStr}} and discussed in Subsec.~\ref{subsec:stp1}), the Fock matrix elements, $ \{\fockel{q}{p} \}$, and the two-electron repulsion integrals, $\{\eriel{rs}{pq}\}$.  
Next, it takes advantage of the antisymmetric permutations of the amplitudes and integrals to obtain more compact expressions.

Finally, as output, \brown{\texttt{ManyBodyBlocks}} obtained in the module \red{\texttt{ManyBodyHamiltonian}} (see Subsec.~\ref{subsec:stp2}) and \brown{\texttt{ManyBodyTerms}}, $\mbt{\ind{h}{1}\ind{h}{2}\cdots \ind{p}{3}\ind{p}{4}\cdots}{\ind{p}{1}\ind{p}{2}\cdots \ind{h}{3}\ind{h}{4}\cdots} \qty[ \{\fockel{q}{p} \},\{\eriel{rs}{pq}\}, \{\at{\ind{h}{1}\ind{h}{2}\cdots}{\ind{p}{1}\ind{p}{2}\cdots}\}  ]$, from \red{\texttt{AntiSymPermut}} are printed.
From these, one can easily generate the EOM-CC equations. 

\subsection{Examples}
\label{subsec:examples}

The software has also the advantage of being able to generate various outputs, including CC energy and amplitude equations, second-quantized strings in normal order, many-body terms of the similarity-transformed Hamiltonian, and blocks of the EOM-CC Hamiltonian matrix in terms of many-body terms.

For example, the input code to obtain IP-EOM-CCSD is
\begin{lstlisting}[language=Mathematica]
	(* Input for IP-EOM-CCSD *)
	
	ClusterOperator={{"1h1p"},{"2h2p"}};
	EOMOperator={{"1h0p"},{"2h1p"}};
	eomccgen[ClusterOperator,EOMOperator]
	
\end{lstlisting}
while the output consists of a list of many-body terms and the EOM-CC matrix in terms of these quantities:
\begin{lstlisting}[extendedchars=true,language=Mathematica,mathescape=true]
    (* Many-body terms *)
    
    $\chi$1$[\![$h3,h1$]\!]$==-F$[\![$h3,h1$]\!]$-F$[\![$h3,v1$]\!]$t1$[\![$h1,v1$]\!]$+t1$[\![$o1,v1$]\!]$ERI$[\![$h3,o1,v1,h1$]\!]$+       
    t1$[\![$h1,v2$]\!]$t1$[\![$o1,v1$]\!]$ERI$[\![$h3,o1,v1,v2$]\!]$+1/2t2$[\![$o1,h1,v1,v2$]\!]$ERI$[\![$h3,o1,v1,v2$]\!]$ 
    $\chi$1$[\![$h4,p3$]\!]$==F$[\![$h4,p3$]\!]$-t1$[\![$o1,v1$]\!]$ERI$[\![$h4,o1,v1,p3$]\!]$
    $\chi$2$[\![$h4,h3,h1,p3$]\!]$==ERI$[\![$h4,h3,h1,p3$]\!]$+t1$[\![$h1,v1$]\!]$ERI$[\![$h4,h3,v1,p3$]\!]$
    $\chi$2$[\![$h3,p1,h1,h2$]\!]$==t1$[\![$o1,p1$]\!]$ERI$[\![$h3,o1,h1,h2$]\!]$-t1$[\![$h2,v1$]\!]$t1$[\![$o1,p1$]\!]$ERI$[\![$h3,o1,v1,h1$]\!]$+
    t1$[\![$h1,v1$]\!]$t1$[\![$o1,p1$]\!]$ERI$[\![$h3,o1,v1,h2$]\!]$+t1$[\![$h1,v1$]\!]$t1$[\![$h2,v2$]\!]$t1$[\![$o1,p1$]\!]$ERI$[\![$h3,o1,v1,v2$]\!]$-
    ERI$[\![$h3,p1,h1,h2$]\!]$+t1$[\![$h2,v1$]\!]$ ERI$[\![$h3,p1,v1,h1$]\!]$-t1$[\![$h1,v1$]\!]$ ERI$[\![$h3,p1,v1,h2$]\!]$-
    t1$[\![$h1,v1$]\!]$t1$[\![$h2,v2$]\!]$ERI$[\![$h3,p1,v1,v2$]\!]$-F$[\![$h3,v1$]\!]$t2$[\![$h1,h2,v1,p1$]\!]$+
    1/2t1$[\![$o1,p1$]\!]$t2$[\![$h1,h2,v1,v2$]\!]$ERI$[\![$h3,o1,v1,v2$]\!]$-1/2ERI$[\![$h3,p1,v1,v2$]\!]$t2$[\![$h1,h2,v1,v2$]\!]$+
    t1$[\![$o1,v1$]\!]$t2$[\![$h1,h2,v2,p1$]\!]$ERI$[\![$h3,o1,v1,v2$]\!]$-t2$[\![$o1,h1,v1,p1$]\!]$ERI$[\![$h3,o1,v1,h2$]\!]$
    +t1$[\![$h2,v1$]\!]$t2$[\![$o1,h1,v2,p1$]\!]$ERI$[\![$h3,o1,v1,v2$]\!]$+t2$[\![$o1,h2,v1,p1$]\!]$ERI$[\![$h3,o1,v1,h1$]\!]$
    -t1$[\![$h1,v1$]\!]$t2$[\![$o1,h2,v2,p1$]\!]$ ERI$[\![$h3,o1,v1,v2$]\!]$ 
    $\chi$1$[\![$p1,p3$]\!]$==F$[\![$p1,p3$]\!]$-F$[\![$o1,p3$]\!]$ t1$[\![$o1,p1$]\!]$-t1$[\![$o1,v1$]\!]$ t1$[\![$o2,p1$]\!]$ ERI$[\![$o1,o2,v1,p3$]\!]$+t1$[\![$o1,v1$]\!]$ ERI$[\![$o1,p1,v1,p3$]\!]$-1/2ERI$[\![$o1,o2,v1,p3$]\!]$t2$[\![$o1,o2,v1,p1$]\!]$
    $\chi$2$[\![$h4,h3,h1,h2$]\!]$==-ERI$[\![$h4,h3,h1,h2$]\!]$+t1$[\![$h2,v1$]\!]$ ERI$[\![$h4,h3,v1,h1$]\!]$-t1$[\![$h1,v1$]\!]$ ERI$[\![$h4,h3,v1,h2$]\!]$-t1$[\![$h1,v1$]\!]$ t1$[\![$h2,v2$]\!]$ ERI$[\![$h4,h3,v1,v2$]\!]$-1/2 ERI$[\![$h4,h3,v1,v2$]\!]$ t2$[\![$h1,h2,v1,v2$]\!]$
    $\chi$2$[\![$h4,p1,h2,p3$]\!]$==t1$[\![$o1,p1$]\!]$ ERI$[\![$h4,o1,h2,p3$]\!]$+t1$[\![$h2,v1$]\!]$t1$[\![$o1,p1$]\!]$ERI$[\![$h4,o1,v1,p3$]\!]$-ERI$[\![$h4,p1,h2,p3$]\!]$
    -t1$[\![$h2,v1$]\!]$ ERI$[\![$h4,p1,v1,p3$]\!]$-ERI$[\![$h4,o1,v1,p3$]\!]$ t2$[\![$o1,h2,v1,p1$]\!]$
    $\chi$3$[\![$h4,h3,p1,h1,h2,p3$]\!]$==ERI$[\![$h4,h3,v1,p3$]\!]$ t2$[\![$h1,h2,v1,p1$]\!]$
    
    (* EOM-CC blocks *)
    
    {{-$\chi$1$[\![$h3,h1$]\!]$}, 
    {$\chi$2$[\![$h4,h3, h1, p3$]\!]$+$\delta_{\texttt{h1,h3}}$$\chi$1$[\![$h4,p3$]\!]$-$\delta_{\texttt{h1,h4}}$$\chi$1$[\![$h3,p3$]\!]$},
    {-$\chi$2$[\![$h3,p1,h2,h1$]\!]$},
    {-$\chi$1$[\![$h4,h2$]\!]$$\delta_{\texttt{h1,h3}}$$\delta_{\texttt{p1,p3}}$+$\chi$1$[\![$h3,h2$]\!]$$\delta_{\texttt{h1,h4}}$$\delta_{\texttt{p1,p3}}$+$\chi$1$[\![$h4,h1$]\!]$$\delta_{\texttt{h2,h3}}$$\delta_{\texttt{p1,p3}}$-$\chi$1$[\![$h3,h1$]\!]$$\delta_{\texttt{h2,h4}}$$\delta_{\texttt{p1,p3}}$-$\chi$1$[\![$p1,p3$]\!]$$\delta_{\texttt{h1,h4}}$$\delta_{\texttt{h2,h3}}$
    +$\chi$1$[\![$p1,p3$]\!]$$\delta_{\texttt{h1,h3}}$$\delta_{\texttt{h2,h4}}$-$\chi$2$[\![$h4,p1,h2,p3$]\!]$$\delta_{\texttt{h1,h3}}$+$\chi$2$[\![$h3,p1,h2,p3$]\!]$$\delta_{\texttt{h1,h4}}$+$\chi$2$[\![$h4,p1, h1,p3$]\!]$$\delta_{\texttt{h2,h3}}$- $\chi$2$[\![$h3,p1,h1,p3$]\!]$$\delta_{\texttt{h2,h4}}$+
    $\chi$2$[\![$h4,h3,h2,h1$]\!]$$\delta_{\texttt{p1,p3}}$+$\chi$3$[\![$h4,h3,p1,h2,h1,p3$]\!]$}}
    
\end{lstlisting}

The input code to generate the CCD equations is
\begin{lstlisting}[language=Mathematica]
	(* Input for CCD *)
	
	ClusterOperator={{"2h2p"}};
	EOMOperator={{"0h0p"}};
	eomccgen[ClusterOperator,EOMOperator]
	
\end{lstlisting}
and it produces the following output:
\begin{lstlisting}[extendedchars=true,language=Mathematica,mathescape=true]
    (* CCD energy *)
    
    {1/4 ERI$[\![$o1,o2,v1,v2$]\!]$ t2$[\![$o1,o2,v1,v2$]\!]$}
    
    (* T2 equations *)
    
    {-ERI$[\![$p2,p1,h1,h2$]\!]$ 
    -F$[\![$p2,v1$]\!]$t2$[\![$h1,h2,v1,p1$]\!]$+F$[\![$p1,v1$]\!]$t2$[\![$h1,h2,v1,p2$]\!]$
    -F$[\![$o1,h2$]\!]$t2$[\![$o1,h1,p2,p1$]\!]$+F$[\![$o1,h1$]\!]$t2$[\![$o1,h2,p2,p1$]\!]$
    -1/2 ERI$[\![$p2,p1,v1,v2$]\!]$t2$[\![$h1,h2,v1,v2$]\!]$ +ERI$[\![$o1,p2,v1,h2$]\!]$t2$[\![$o1,h1,v1,p1$]\!]$
    -ERI$[\![$o1,p1,v1,h2$]\!]$t2$[\![$o1,h1,v1,p2$]\!]$-ERI$[\![$o1,p2,v1,h1$]\!]$t2$[\![$o1,h2,v1,p1$]\!]$
    +ERI$[\![$o1,p1,v1,h1$]\!]$t2$[\![$o1,h2,v1,p2$]\!]$-1/2ERI$[\![$o1,o2,h1,h2$]\!]$t2$[\![$o1,o2,p2,p1$]\!]$
    -1/4ERI$[\![$o1,o2,v1,v2$]\!]$t2$[\![$h1,h2,v1,v2$]\!]$t2$[\![$o1,o2,p2,p1$]\!]$
    -1/2ERI$[\![$$[\![$o1,o2,v1,v2$]\!]$t2$[\![$h1,h2,v2,p2$]\!]$t2$[\![$o1,o2,v1,p1$]\!]$
    +1/2ERI$[\![$o1,o2,v1,v2$]\!]$t2$[\![$h1,h2,v2,p1$]\!]$t2$[\![$o1,o2,v1,p2$]\!]$
    -1/2ERI$[\![$o1,o2,v1,v2$]\!]$t2$[\![$o1,h2,v1,v2$]\!]$t2$[\![$o2,h1,p2,p1$]\!]$
    +1/2ERI$[\![$o1,o2,v1,v2$]\!]$t2$[\![$o1,h1,v1,v2$]\!]$t2$[\![$o2,h2,p2,p1$]\!]$
    +ERI$[\![$o1,o2,v1,v2$]\!]$t2$[\![$o1,h2,v1,p2$]\!]$t2$[\![$o2,h1,v2,p1$]\!]$
    -ERI$[\![$o1,o2,v1,v2$]\!]$t2$[\![$o1,h1,v1,p2$]\!]$t2$[\![$o2,h2,v2,p1$]\!]$} 
    
\end{lstlisting}

In the following example, we compute the block \texttt{\{2,1\}} of the IP-EOM-CCSD matrix.
The corresponding input is
\begin{lstlisting}[language=Mathematica]
	(* Input for IP-EOM-CCSD *)
	
	ClusterOperator={{"1h0p"},{"2h1p"}};
	EOMOperator={{"1h1p"},{"2h2p"}};
	EOMBlock={2,1};
	eomccgen[ClusterOperator,EOMOperator,EOMBlock]
	
\end{lstlisting}
and the output is
\begin{lstlisting}[language=Mathematica,mathescape=true]
    (* Many body terms *)
    
    $\chi$1$[\![$h4,p3$]\!]$==F$[\![$h4,p3$]\!]$-ERI$[\![$h4,o1,v1,p3$]\!]$t1$[\![$o1,v1$]\!]$
    $\chi$2$[\![$h4,h3,h1,p3$]\!]$==ERI$[\![$h4,h3,h1,p3$]\!]$-ERI$[\![$h4,h3,v1,p3$]\!]$t1$[\![$h1,v1$]\!]$
    
    (* EOM-CC blocks *)
    
    {$\chi$1$[\![$h4,p3$]\!]$$\delta_{\texttt{h1,h3}}$-$\chi$1$[\![$h3,p3$]\!]$$\delta_{\texttt{h1,h4}}$+$\chi$2$[\![$h4,h3,v1,p3$]\!]$}

\end{lstlisting}
\alert{It is also possible to contract the EOM-CC Hamiltonian and the right or left eigenvectors [see Eqs.~\eqref{eq:eomccmf} and \eqref{eq:eomccmfLeft}]. These quantities are useful when one relies on the Davidson diagonalization method \cite{davidson1975} or related techniques. For example, the input to obtain the left-hand eigenvalue equation for IP-EOM-CCSD is the following:}
\begin{lstlisting}[language=Mathematica]
	(* Input for IP-EOM-CCSD *)
	
	ClusterOperator={{"1h0p"},{"2h1p"}};
	EOMOperator={{"1h1p"},{"2h2p"}};
	LeftRigthEOM={"L"};
	eomccgen[ClusterOperator,EOMOperator,LeftRigthEOM]
	
\end{lstlisting}
\alert{The output is very similar to the first example with the difference that, in} \brown{\texttt{ManyBodyBlocks}}\alert{, the Kronecker deltas are applied to the many-body terms as well as to the left-hand EOM-CC amplitudes.}

The $\mat$ code described in this paper is publicly available in a dedicated repository on \textsc{github} with the name \red{\texttt{eomccgen}}. \cite{rauleomccgen2023} 
In particular, the repository contains examples of scripts and data files that can be easily adapted for particular research projects. Additionally, the repository includes another notebook, called \red{\texttt{eomccnum}}, where the equations obtained in \red{\texttt{eomccgen}} can be implemented and numerically tested for small atoms and molecules. \red{\texttt{eomccnum}} comes with pre-installed examples that can serve as a guide for implementing the equations obtained in \red{\texttt{eomccgen}}. An example of the water molecule in the STO-3G basis can be found in the repository.

\section{Conclusion}
\label{sec:conclusion}

The present paper discusses the development of a code generator for equation-of-motion coupled-cluster (EOM-CC) methods, a class of many-body quantum chemistry methods known for their accuracy in predicting excitation energies, ionization potentials, and electron affinities in molecular systems.
Because their implementation can be complex and time-consuming, we develop an easy-to-use code generator that automates the process of deriving and implementing the EOM-CC equations, reducing the potential for human error.

We begin by discussing the second-quantization formalism, a practical and modern mathematical language used to describe many-body quantum systems in terms of creation and annihilation operators. 
We then discuss the development of our code generator for EOM-CC methods. 
Our approach builds on previous work by using second-quantized strings to automate the derivation and implementation of EOM-CC equations. 
However, we introduce several new features that make the code generator more efficient and flexible. 
For example, we rely on a symbolic algebra software package, \mat, to generate these equations that can be easily read by humans and machines.
The paper provides a detailed description of each step of the algorithm used in the code generator. 
We also describe how the code generator can be used to calculate excitation energies, ionization potentials, and electron affinities, by providing several concrete examples.

Although many improvements are still needed to generate a production-level code, especially in the definition of adequate intermediates to ensure the right computational scaling of the methods, we hope that the present work illustrates nicely the capabilities of \mat-based code generator in the context of CC theory.


\acknowledgments{
The authors thank Yann Damour and Antoine Marie for useful comments on the manuscript.
This project has received funding from the European Research Council (ERC) under the European Union's Horizon 2020 research and innovation programme (Grant agreement No.~863481).}

\bibliography{equationgenerator}

\begin{thebibliography}{80}%
\makeatletter
\providecommand \@ifxundefined [1]{%
 \@ifx{#1\undefined}
}%
\providecommand \@ifnum [1]{%
 \ifnum #1\expandafter \@firstoftwo
 \else \expandafter \@secondoftwo
 \fi
}%
\providecommand \@ifx [1]{%
 \ifx #1\expandafter \@firstoftwo
 \else \expandafter \@secondoftwo
 \fi
}%
\providecommand \natexlab [1]{#1}%
\providecommand \enquote  [1]{``#1''}%
\providecommand \bibnamefont  [1]{#1}%
\providecommand \bibfnamefont [1]{#1}%
\providecommand \citenamefont [1]{#1}%
\providecommand \href@noop [0]{\@secondoftwo}%
\providecommand \href [0]{\begingroup \@sanitize@url \@href}%
\providecommand \@href[1]{\@@startlink{#1}\@@href}%
\providecommand \@@href[1]{\endgroup#1\@@endlink}%
\providecommand \@sanitize@url [0]{\catcode `\\12\catcode `\$12\catcode
  `\&12\catcode `\#12\catcode `\^12\catcode `\_12\catcode `\%12\relax}%
\providecommand \@@startlink[1]{}%
\providecommand \@@endlink[0]{}%
\providecommand \url  [0]{\begingroup\@sanitize@url \@url }%
\providecommand \@url [1]{\endgroup\@href {#1}{\urlprefix }}%
\providecommand \urlprefix  [0]{URL }%
\providecommand \Eprint [0]{\href }%
\providecommand \doibase [0]{https://doi.org/}%
\providecommand \selectlanguage [0]{\@gobble}%
\providecommand \bibinfo  [0]{\@secondoftwo}%
\providecommand \bibfield  [0]{\@secondoftwo}%
\providecommand \translation [1]{[#1]}%
\providecommand \BibitemOpen [0]{}%
\providecommand \bibitemStop [0]{}%
\providecommand \bibitemNoStop [0]{.\EOS\space}%
\providecommand \EOS [0]{\spacefactor3000\relax}%
\providecommand \BibitemShut  [1]{\csname bibitem#1\endcsname}%
\let\auto@bib@innerbib\@empty
\bibitem [{\citenamefont {Dirac}\ and\ \citenamefont {Bohr}(1927)}]{dirac1927}%
  \BibitemOpen
  \bibfield  {author} {\bibinfo {author} {\bibfnamefont {P.~A.~M.}\
  \bibnamefont {Dirac}}\ and\ \bibinfo {author} {\bibfnamefont {N.~H.~D.}\
  \bibnamefont {Bohr}},\ }\bibfield  {title} {\enquote {\bibinfo {title} {The
  quantum theory of the emission and absorption of radiation},}\ }\href
  {https://doi.org/10.1098/rspa.1927.0039} {\bibfield  {journal} {\bibinfo
  {journal} {Proceedings of the Royal Society of London. Series A, Containing
  Papers of a Mathematical and Physical Character}\ }\textbf {\bibinfo {volume}
  {114}},\ \bibinfo {pages} {243--265} (\bibinfo {year} {1927})}\BibitemShut
  {NoStop}%
\bibitem [{\citenamefont {Shavitt}\ and\ \citenamefont
  {Bartlett}(2009)}]{shavitt2009}%
  \BibitemOpen
  \bibfield  {author} {\bibinfo {author} {\bibfnamefont {I.}~\bibnamefont
  {Shavitt}}\ and\ \bibinfo {author} {\bibfnamefont {R.~J.}\ \bibnamefont
  {Bartlett}},\ }\href {https://doi.org/10.1017/CBO9780511596834} {\emph
  {\bibinfo {title} {Many-Body Methods in Chemistry and Physics: MBPT and
  Coupled-Cluster Theory}}},\ Cambridge Molecular Science\ (\bibinfo
  {publisher} {Cambridge University Press},\ \bibinfo {year}
  {2009})\BibitemShut {NoStop}%
\bibitem [{\citenamefont {Wick}(1950)}]{wick1950}%
  \BibitemOpen
  \bibfield  {author} {\bibinfo {author} {\bibfnamefont {G.~C.}\ \bibnamefont
  {Wick}},\ }\bibfield  {title} {\enquote {\bibinfo {title} {The evaluation of
  the collision matrix},}\ }\href {https://doi.org/10.1103/PhysRev.80.268}
  {\bibfield  {journal} {\bibinfo  {journal} {Phys. Rev.}\ }\textbf {\bibinfo
  {volume} {80}},\ \bibinfo {pages} {268--272} (\bibinfo {year}
  {1950})}\BibitemShut {NoStop}%
\bibitem [{\citenamefont {Matthews}(2018)}]{matthews2018}%
  \BibitemOpen
  \bibfield  {author} {\bibinfo {author} {\bibfnamefont {D.~A.}\ \bibnamefont
  {Matthews}},\ }\bibfield  {title} {\enquote {\bibinfo {title}
  {High-performance tensor contraction without transposition},}\ }\href
  {https://doi.org/10.1137/16M108968X} {\bibfield  {journal} {\bibinfo
  {journal} {SIAM J. Sci. Comput.}\ }\textbf {\bibinfo {volume} {40}},\
  \bibinfo {pages} {C1--C24} (\bibinfo {year} {2018})}\BibitemShut {NoStop}%
\bibitem [{\citenamefont {Abdelfattah}\ \emph {et~al.}(2016)\citenamefont
  {Abdelfattah}, \citenamefont {Baboulin}, \citenamefont {Dobrev},
  \citenamefont {Dongarra}, \citenamefont {Earl}, \citenamefont {Falcou},
  \citenamefont {Haidar}, \citenamefont {Karlin}, \citenamefont {Kolev},
  \citenamefont {Masliah},\ and\ \citenamefont {Tomov}}]{abdelfattah2016}%
  \BibitemOpen
  \bibfield  {author} {\bibinfo {author} {\bibfnamefont {A.}~\bibnamefont
  {Abdelfattah}}, \bibinfo {author} {\bibfnamefont {M.}~\bibnamefont
  {Baboulin}}, \bibinfo {author} {\bibfnamefont {V.}~\bibnamefont {Dobrev}},
  \bibinfo {author} {\bibfnamefont {J.}~\bibnamefont {Dongarra}}, \bibinfo
  {author} {\bibfnamefont {C.}~\bibnamefont {Earl}}, \bibinfo {author}
  {\bibfnamefont {J.}~\bibnamefont {Falcou}}, \bibinfo {author} {\bibfnamefont
  {A.}~\bibnamefont {Haidar}}, \bibinfo {author} {\bibfnamefont
  {I.}~\bibnamefont {Karlin}}, \bibinfo {author} {\bibfnamefont
  {T.}~\bibnamefont {Kolev}}, \bibinfo {author} {\bibfnamefont
  {I.}~\bibnamefont {Masliah}},\ and\ \bibinfo {author} {\bibfnamefont
  {S.}~\bibnamefont {Tomov}},\ }\bibfield  {title} {\enquote {\bibinfo {title}
  {High-performance tensor contractions for gpus},}\ }\href
  {https://doi.org/https://doi.org/10.1016/j.procs.2016.05.302} {\bibfield
  {journal} {\bibinfo  {journal} {Procedia Computer Science}\ }\textbf
  {\bibinfo {volume} {80}},\ \bibinfo {pages} {108--118} (\bibinfo {year}
  {2016})},\ \bibinfo {note} {international Conference on Computational Science
  2016, ICCS 2016, 6-8 June 2016, San Diego, California, USA}\BibitemShut
  {NoStop}%
\bibitem [{\citenamefont {Springer}, \citenamefont {Su},\ and\ \citenamefont
  {Bientinesi}(2017)}]{paul2017}%
  \BibitemOpen
  \bibfield  {author} {\bibinfo {author} {\bibfnamefont {P.}~\bibnamefont
  {Springer}}, \bibinfo {author} {\bibfnamefont {T.}~\bibnamefont {Su}},\ and\
  \bibinfo {author} {\bibfnamefont {P.}~\bibnamefont {Bientinesi}},\ }\bibfield
   {title} {\enquote {\bibinfo {title} {Hptt: A high-performance tensor
  transposition c++ library},}\ }in\ \href
  {https://doi.org/10.1145/3091966.3091968} {\emph {\bibinfo {booktitle}
  {Proceedings of the 4th ACM SIGPLAN International Workshop on Libraries,
  Languages, and Compilers for Array Programming}}},\ \bibinfo {series and
  number} {ARRAY 2017}\ (\bibinfo  {publisher} {Association for Computing
  Machinery},\ \bibinfo {address} {New York, NY, USA},\ \bibinfo {year}
  {2017})\ pp.\ \bibinfo {pages} {56--62}\BibitemShut {NoStop}%
\bibitem [{\citenamefont {Calvin}, \citenamefont {Lewis},\ and\ \citenamefont
  {Valeev}(2015)}]{justus2015}%
  \BibitemOpen
  \bibfield  {author} {\bibinfo {author} {\bibfnamefont {J.~A.}\ \bibnamefont
  {Calvin}}, \bibinfo {author} {\bibfnamefont {C.~A.}\ \bibnamefont {Lewis}},\
  and\ \bibinfo {author} {\bibfnamefont {E.~F.}\ \bibnamefont {Valeev}},\
  }\bibfield  {title} {\enquote {\bibinfo {title} {Scalable task-based
  algorithm for multiplication of block-rank-sparse matrices},}\ }in\ \href
  {https://doi.org/10.1145/2833179.2833186} {\emph {\bibinfo {booktitle}
  {Proceedings of the 5th Workshop on Irregular Applications: Architectures and
  Algorithms}}},\ \bibinfo {series and number} {IA3 '15}\ (\bibinfo
  {publisher} {Association for Computing Machinery},\ \bibinfo {address} {New
  York, NY, USA},\ \bibinfo {year} {2015})\BibitemShut {NoStop}%
\bibitem [{\citenamefont {Solomonik}\ \emph {et~al.}(2014)\citenamefont
  {Solomonik}, \citenamefont {Matthews}, \citenamefont {Hammond},\ and\
  \citenamefont {Demmel}}]{solomonik2014}%
  \BibitemOpen
  \bibfield  {author} {\bibinfo {author} {\bibfnamefont {E.}~\bibnamefont
  {Solomonik}}, \bibinfo {author} {\bibfnamefont {D.}~\bibnamefont {Matthews}},
  \bibinfo {author} {\bibfnamefont {J.~F.}\ \bibnamefont {Hammond},
  \bibfnamefont {Jeff R~andStanton}},\ and\ \bibinfo {author} {\bibfnamefont
  {J.}~\bibnamefont {Demmel}},\ }\bibfield  {title} {\enquote {\bibinfo {title}
  {A massively parallel tensor contraction framework for
  coupled-clustercomputations},}\ }\href@noop {} {\bibfield  {journal}
  {\bibinfo  {journal} {Journal of Parallel and Distributed Computing}\
  }\textbf {\bibinfo {volume} {74}},\ \bibinfo {pages} {3176--3190} (\bibinfo
  {year} {2014})}\BibitemShut {NoStop}%
\bibitem [{\citenamefont {Herault}\ \emph {et~al.}(2020)\citenamefont
  {Herault}, \citenamefont {Robert}, \citenamefont {Bosilca}, \citenamefont
  {Harrison}, \citenamefont {Lewis}, \citenamefont {Valeev},\ and\
  \citenamefont {Dongarra}}]{herault2020}%
  \BibitemOpen
  \bibfield  {author} {\bibinfo {author} {\bibfnamefont {T.}~\bibnamefont
  {Herault}}, \bibinfo {author} {\bibfnamefont {Y.}~\bibnamefont {Robert}},
  \bibinfo {author} {\bibfnamefont {G.}~\bibnamefont {Bosilca}}, \bibinfo
  {author} {\bibfnamefont {R.~J.}\ \bibnamefont {Harrison}}, \bibinfo {author}
  {\bibfnamefont {C.~A.}\ \bibnamefont {Lewis}}, \bibinfo {author}
  {\bibfnamefont {E.~F.}\ \bibnamefont {Valeev}},\ and\ \bibinfo {author}
  {\bibfnamefont {J.~J.}\ \bibnamefont {Dongarra}},\ }\href@noop {} {\enquote
  {\bibinfo {title} {{Distributed-memory multi-GPU block-sparse tensor
  contraction for electronic structure (revised version)}},}\ }\bibinfo {type}
  {Research Report}\ \bibinfo {number} {RR-9365}\ (\bibinfo  {institution}
  {{Inria - Research Centre Grenoble -- Rh{\^o}ne-Alpes}},\ \bibinfo {year}
  {2020})\BibitemShut {NoStop}%
\bibitem [{\citenamefont {Psarras}\ \emph {et~al.}(2021)\citenamefont
  {Psarras}, \citenamefont {Karlsson}, \citenamefont {Li},\ and\ \citenamefont
  {Bientinesi}}]{psarras2021}%
  \BibitemOpen
  \bibfield  {author} {\bibinfo {author} {\bibfnamefont {C.}~\bibnamefont
  {Psarras}}, \bibinfo {author} {\bibfnamefont {L.}~\bibnamefont {Karlsson}},
  \bibinfo {author} {\bibfnamefont {J.}~\bibnamefont {Li}},\ and\ \bibinfo
  {author} {\bibfnamefont {P.}~\bibnamefont {Bientinesi}},\ }\href
  {https://doi.org/10.48550/ARXIV.2103.13756} {\enquote {\bibinfo {title} {The
  landscape of software for tensor computations},}\ } (\bibinfo {year}
  {2021})\BibitemShut {NoStop}%
\bibitem [{\citenamefont {Coester}(1958)}]{coester1957}%
  \BibitemOpen
  \bibfield  {author} {\bibinfo {author} {\bibfnamefont {F.}~\bibnamefont
  {Coester}},\ }\bibfield  {title} {\enquote {\bibinfo {title} {Bound states of
  a many-particle system},}\ }\href
  {https://doi.org/https://doi.org/10.1016/0029-5582(58)90280-3} {\bibfield
  {journal} {\bibinfo  {journal} {Nuc. Phys.}\ }\textbf {\bibinfo {volume}
  {7}},\ \bibinfo {pages} {421--424} (\bibinfo {year} {1958})}\BibitemShut
  {NoStop}%
\bibitem [{\citenamefont {Coester}\ and\ \citenamefont
  {K{\"u}mmel}(1960)}]{coester1960}%
  \BibitemOpen
  \bibfield  {author} {\bibinfo {author} {\bibfnamefont {F.}~\bibnamefont
  {Coester}}\ and\ \bibinfo {author} {\bibfnamefont {H.}~\bibnamefont
  {K{\"u}mmel}},\ }\bibfield  {title} {\enquote {\bibinfo {title} {Short-range
  correlations in nuclear wave functions},}\ }\href
  {https://doi.org/https://doi.org/10.1016/0029-5582(60)90140-1} {\bibfield
  {journal} {\bibinfo  {journal} {Nuc. Phys..}\ }\textbf {\bibinfo {volume}
  {17}},\ \bibinfo {pages} {477--485} (\bibinfo {year} {1960})}\BibitemShut
  {NoStop}%
\bibitem [{\citenamefont {{\v C}{\'\i}{\v z}ek}(1966)}]{cizek1966}%
  \BibitemOpen
  \bibfield  {author} {\bibinfo {author} {\bibfnamefont {J.}~\bibnamefont {{\v
  C}{\'\i}{\v z}ek}},\ }\bibfield  {title} {\enquote {\bibinfo {title} {On the
  correlation problem in atomic and molecular systems. calculation of
  wavefunction components in ursell‐type expansion using quantum‐field
  theoretical methods},}\ }\href {https://doi.org/10.1063/1.1727484} {\bibfield
   {journal} {\bibinfo  {journal} {J. Chem. Phys.}\ }\textbf {\bibinfo {volume}
  {45}},\ \bibinfo {pages} {4256--4266} (\bibinfo {year} {1966})}\BibitemShut
  {NoStop}%
\bibitem [{\citenamefont {Crawford}\ and\ \citenamefont
  {Schaefer~III}(2000)}]{crawford2000}%
  \BibitemOpen
  \bibfield  {author} {\bibinfo {author} {\bibfnamefont {T.~D.}\ \bibnamefont
  {Crawford}}\ and\ \bibinfo {author} {\bibfnamefont {H.~F.}\ \bibnamefont
  {Schaefer~III}},\ }\enquote {\bibinfo {title} {An introduction to coupled
  cluster theory for computational chemists},}\ in\ \href
  {https://doi.org/https://doi.org/10.1002/9780470125915.ch2} {\emph {\bibinfo
  {booktitle} {Reviews in Computational Chemistry}}}\ (\bibinfo  {publisher}
  {John Wiley and Sons, Ltd},\ \bibinfo {year} {2000})\ pp.\ \bibinfo {pages}
  {33--136}\BibitemShut {NoStop}%
\bibitem [{\citenamefont {Janssen}\ and\ \citenamefont
  {Schaefer}(1991)}]{janssen1991}%
  \BibitemOpen
  \bibfield  {author} {\bibinfo {author} {\bibfnamefont {C.~L.}\ \bibnamefont
  {Janssen}}\ and\ \bibinfo {author} {\bibfnamefont {H.~F.}\ \bibnamefont
  {Schaefer}},\ }\bibfield  {title} {\enquote {\bibinfo {title} {The automated
  solution of second quantization equations with applications to the coupled
  cluster approach},}\ }\href
  {https://doi.org/https://doi.org/10.1007/BF01113327} {\bibfield  {journal}
  {\bibinfo  {journal} {Theor. Chim. Acta}\ }\textbf {\bibinfo {volume} {79}},\
  \bibinfo {pages} {1--42} (\bibinfo {year} {1991})}\BibitemShut {NoStop}%
\bibitem [{\citenamefont {Li}\ and\ \citenamefont {Paldus}(1994)}]{li1994}%
  \BibitemOpen
  \bibfield  {author} {\bibinfo {author} {\bibfnamefont {X.}~\bibnamefont
  {Li}}\ and\ \bibinfo {author} {\bibfnamefont {J.}~\bibnamefont {Paldus}},\
  }\bibfield  {title} {\enquote {\bibinfo {title} {Automation of the
  implementation of spin‐adapted open‐shell coupled‐cluster theories
  relying on the unitary group formalism},}\ }\href
  {https://doi.org/10.1063/1.468074} {\bibfield  {journal} {\bibinfo  {journal}
  {J. Chem. Phys.}\ }\textbf {\bibinfo {volume} {101}},\ \bibinfo {pages}
  {8812--8826} (\bibinfo {year} {1994})}\BibitemShut {NoStop}%
\bibitem [{\citenamefont {K{\'a}llay}\ and\ \citenamefont
  {Surj{\'a}n}(2001)}]{kallay2001}%
  \BibitemOpen
  \bibfield  {author} {\bibinfo {author} {\bibfnamefont {M.}~\bibnamefont
  {K{\'a}llay}}\ and\ \bibinfo {author} {\bibfnamefont {P.~R.}\ \bibnamefont
  {Surj{\'a}n}},\ }\bibfield  {title} {\enquote {\bibinfo {title} {Higher
  excitations in coupled-cluster theory},}\ }\href
  {https://doi.org/10.1063/1.1383290} {\bibfield  {journal} {\bibinfo
  {journal} {J. Chem. Phys.}\ }\textbf {\bibinfo {volume} {115}},\ \bibinfo
  {pages} {2945--2954} (\bibinfo {year} {2001})}\BibitemShut {NoStop}%
\bibitem [{\citenamefont {Das}, \citenamefont {Mukherjee},\ and\ \citenamefont
  {K{\'a}llay}(2010)}]{kallay2010}%
  \BibitemOpen
  \bibfield  {author} {\bibinfo {author} {\bibfnamefont {S.}~\bibnamefont
  {Das}}, \bibinfo {author} {\bibfnamefont {D.}~\bibnamefont {Mukherjee}},\
  and\ \bibinfo {author} {\bibfnamefont {M.}~\bibnamefont {K{\'a}llay}},\
  }\bibfield  {title} {\enquote {\bibinfo {title} {Full implementation and
  benchmark studies of mukherjee's state-specific multireference
  coupled-cluster ansatz},}\ }\href {https://doi.org/10.1063/1.3310288}
  {\bibfield  {journal} {\bibinfo  {journal} {J. Chem. Phys.}\ }\textbf
  {\bibinfo {volume} {132}},\ \bibinfo {pages} {074103} (\bibinfo {year}
  {2010})}\BibitemShut {NoStop}%
\bibitem [{\citenamefont {K{\'a}llay}\ and\ \citenamefont
  {Gauss}(2004)}]{kallay2004}%
  \BibitemOpen
  \bibfield  {author} {\bibinfo {author} {\bibfnamefont {M.}~\bibnamefont
  {K{\'a}llay}}\ and\ \bibinfo {author} {\bibfnamefont {J.}~\bibnamefont
  {Gauss}},\ }\bibfield  {title} {\enquote {\bibinfo {title} {Calculation of
  excited-state properties using general coupled-cluster and
  configuration-interaction models},}\ }\href
  {https://doi.org/10.1063/1.1805494} {\bibfield  {journal} {\bibinfo
  {journal} {J. Chem. Phys.}\ }\textbf {\bibinfo {volume} {121}},\ \bibinfo
  {pages} {9257--9269} (\bibinfo {year} {2004})}\BibitemShut {NoStop}%
\bibitem [{\citenamefont {K{\'a}llay}\ and\ \citenamefont
  {Gauss}(2005)}]{kallay2005}%
  \BibitemOpen
  \bibfield  {author} {\bibinfo {author} {\bibfnamefont {M.}~\bibnamefont
  {K{\'a}llay}}\ and\ \bibinfo {author} {\bibfnamefont {J.}~\bibnamefont
  {Gauss}},\ }\bibfield  {title} {\enquote {\bibinfo {title} {Approximate
  treatment of higher excitations in coupled-cluster theory},}\ }\href
  {https://doi.org/10.1063/1.2121589} {\bibfield  {journal} {\bibinfo
  {journal} {J. Chem. Phys.}\ }\textbf {\bibinfo {volume} {123}},\ \bibinfo
  {pages} {214105} (\bibinfo {year} {2005})}\BibitemShut {NoStop}%
\bibitem [{\citenamefont {Hirata}(2003)}]{hirata2003}%
  \BibitemOpen
  \bibfield  {author} {\bibinfo {author} {\bibfnamefont {S.}~\bibnamefont
  {Hirata}},\ }\bibfield  {title} {\enquote {\bibinfo {title} {Tensor
  contraction engine:  abstraction and automated parallel implementation of
  configuration-interaction, coupled-cluster, and many-body perturbation
  theories},}\ }\href {https://doi.org/10.1021/jp034596z} {\bibfield  {journal}
  {\bibinfo  {journal} {J. Phys. Chem. A}\ }\textbf {\bibinfo {volume} {107}},\
  \bibinfo {pages} {9887--9897} (\bibinfo {year} {2003})}\BibitemShut {NoStop}%
\bibitem [{\citenamefont {Hirata}(2004)}]{hirata2004}%
  \BibitemOpen
  \bibfield  {author} {\bibinfo {author} {\bibfnamefont {S.}~\bibnamefont
  {Hirata}},\ }\bibfield  {title} {\enquote {\bibinfo {title} {Higher-order
  equation-of-motion coupled-cluster methods},}\ }\href
  {https://doi.org/10.1063/1.1753556} {\bibfield  {journal} {\bibinfo
  {journal} {J. Chem. Phys.}\ }\textbf {\bibinfo {volume} {121}},\ \bibinfo
  {pages} {51--59} (\bibinfo {year} {2004})}\BibitemShut {NoStop}%
\bibitem [{\citenamefont {Auer}\ \emph {et~al.}(2006)\citenamefont {Auer},
  \citenamefont {Baumgartner}, \citenamefont {Bernholdt}, \citenamefont
  {Bibireata}, \citenamefont {Choppella}, \citenamefont {Cociorva},
  \citenamefont {Gao}, \citenamefont {Harrison}, \citenamefont
  {Krishnamoorthy}, \citenamefont {Krishnan}, \citenamefont {Lam},
  \citenamefont {Lu}, \citenamefont {Nooijen}, \citenamefont {Pitzer},
  \citenamefont {Ramanujam}, \citenamefont {Sadayappan},\ and\ \citenamefont
  {Sibiryakov}}]{alexander2006}%
  \BibitemOpen
  \bibfield  {author} {\bibinfo {author} {\bibfnamefont {A.~A.}\ \bibnamefont
  {Auer}}, \bibinfo {author} {\bibfnamefont {G.}~\bibnamefont {Baumgartner}},
  \bibinfo {author} {\bibfnamefont {D.~E.}\ \bibnamefont {Bernholdt}}, \bibinfo
  {author} {\bibfnamefont {A.}~\bibnamefont {Bibireata}}, \bibinfo {author}
  {\bibfnamefont {V.}~\bibnamefont {Choppella}}, \bibinfo {author}
  {\bibfnamefont {D.}~\bibnamefont {Cociorva}}, \bibinfo {author}
  {\bibfnamefont {X.}~\bibnamefont {Gao}}, \bibinfo {author} {\bibfnamefont
  {R.}~\bibnamefont {Harrison}}, \bibinfo {author} {\bibfnamefont
  {S.}~\bibnamefont {Krishnamoorthy}}, \bibinfo {author} {\bibfnamefont
  {S.}~\bibnamefont {Krishnan}}, \bibinfo {author} {\bibfnamefont {C.-C.}\
  \bibnamefont {Lam}}, \bibinfo {author} {\bibfnamefont {Q.}~\bibnamefont
  {Lu}}, \bibinfo {author} {\bibfnamefont {M.}~\bibnamefont {Nooijen}},
  \bibinfo {author} {\bibfnamefont {R.}~\bibnamefont {Pitzer}}, \bibinfo
  {author} {\bibfnamefont {J.}~\bibnamefont {Ramanujam}}, \bibinfo {author}
  {\bibfnamefont {P.}~\bibnamefont {Sadayappan}},\ and\ \bibinfo {author}
  {\bibfnamefont {A.}~\bibnamefont {Sibiryakov}},\ }\bibfield  {title}
  {\enquote {\bibinfo {title} {Automatic code generation for many-body
  electronic structure methods: the tensor contraction engine},}\ }\href
  {https://doi.org/10.1080/00268970500275780} {\bibfield  {journal} {\bibinfo
  {journal} {Mol. Phys.}\ }\textbf {\bibinfo {volume} {104}},\ \bibinfo {pages}
  {211--228} (\bibinfo {year} {2006})}\BibitemShut {NoStop}%
\bibitem [{\citenamefont {Kamiya}\ and\ \citenamefont
  {Hirata}(2006{\natexlab{a}})}]{hirata2006}%
  \BibitemOpen
  \bibfield  {author} {\bibinfo {author} {\bibfnamefont {M.}~\bibnamefont
  {Kamiya}}\ and\ \bibinfo {author} {\bibfnamefont {S.}~\bibnamefont
  {Hirata}},\ }\bibfield  {title} {\enquote {\bibinfo {title} {Higher-order
  equation-of-motion coupled-cluster methods for ionization processes},}\
  }\href {https://doi.org/10.1063/1.2244570} {\bibfield  {journal} {\bibinfo
  {journal} {J. Chem. Phys.}\ }\textbf {\bibinfo {volume} {125}},\ \bibinfo
  {pages} {074111} (\bibinfo {year} {2006}{\natexlab{a}})}\BibitemShut
  {NoStop}%
\bibitem [{\citenamefont {Stanton}\ and\ \citenamefont
  {Bartlett}(1993)}]{stanton1993}%
  \BibitemOpen
  \bibfield  {author} {\bibinfo {author} {\bibfnamefont {J.~F.}\ \bibnamefont
  {Stanton}}\ and\ \bibinfo {author} {\bibfnamefont {R.~J.}\ \bibnamefont
  {Bartlett}},\ }\bibfield  {title} {\enquote {\bibinfo {title} {The equation
  of motion coupled‐cluster method. a systematic biorthogonal approach to
  molecular excitation energies, transition probabilities, and excited state
  properties},}\ }\href {https://doi.org/10.1063/1.464746} {\bibfield
  {journal} {\bibinfo  {journal} {J. Chem. Phys.}\ }\textbf {\bibinfo {volume}
  {98}},\ \bibinfo {pages} {7029--7039} (\bibinfo {year} {1993})}\BibitemShut
  {NoStop}%
\bibitem [{\citenamefont {Bartlett}(2012)}]{barlett2012}%
  \BibitemOpen
  \bibfield  {author} {\bibinfo {author} {\bibfnamefont {R.~J.}\ \bibnamefont
  {Bartlett}},\ }\bibfield  {title} {\enquote {\bibinfo {title}
  {Coupled-cluster theory and its equation-of-motion extensions},}\ }\href
  {https://doi.org/https://doi.org/10.1002/wcms.76} {\bibfield  {journal}
  {\bibinfo  {journal} {WIRES Comput. Mol. Sci.}\ }\textbf {\bibinfo {volume}
  {2}},\ \bibinfo {pages} {126--138} (\bibinfo {year} {2012})}\BibitemShut
  {NoStop}%
\bibitem [{\citenamefont {Krylov}(2008)}]{krylov2008}%
  \BibitemOpen
  \bibfield  {author} {\bibinfo {author} {\bibfnamefont {A.~I.}\ \bibnamefont
  {Krylov}},\ }\bibfield  {title} {\enquote {\bibinfo {title}
  {Equation-of-motion coupled-cluster methods for open-shell and electronically
  excited species: The hitchhiker's guide to fock space},}\ }\href
  {https://doi.org/10.1146/annurev.physchem.59.032607.093602} {\bibfield
  {journal} {\bibinfo  {journal} {Annu. Rev. Phys. Chem.}\ }\textbf {\bibinfo
  {volume} {59}},\ \bibinfo {pages} {433--462} (\bibinfo {year} {2008})},\
  \bibinfo {note} {pMID: 18173379}\BibitemShut {NoStop}%
\bibitem [{\citenamefont {Musia{\l}}(2020)}]{musial2020}%
  \BibitemOpen
  \bibfield  {author} {\bibinfo {author} {\bibfnamefont {M.}~\bibnamefont
  {Musia{\l}}},\ }\enquote {\bibinfo {title} {Equation-of-motion
  coupled-cluster models},}\ in\ \href
  {https://doi.org/https://doi.org/10.1002/9781119417774.ch4} {\emph {\bibinfo
  {booktitle} {Quantum Chemistry and Dynamics of Excited States}}}\ (\bibinfo
  {publisher} {John Wiley and Sons, Ltd},\ \bibinfo {year} {2020})\
  Chap.~\bibinfo {chapter} {4}, pp.\ \bibinfo {pages} {77--108}\BibitemShut
  {NoStop}%
\bibitem [{\citenamefont {Emrich}(1981)}]{emrich1981}%
  \BibitemOpen
  \bibfield  {author} {\bibinfo {author} {\bibfnamefont {K.}~\bibnamefont
  {Emrich}},\ }\bibfield  {title} {\enquote {\bibinfo {title} {An extension of
  the coupled cluster formalism to excited states (i)},}\ }\href
  {https://doi.org/https://doi.org/10.1016/0375-9474(81)90179-2} {\bibfield
  {journal} {\bibinfo  {journal} {Nuc. Phys. A}\ }\textbf {\bibinfo {volume}
  {351}},\ \bibinfo {pages} {379--396} (\bibinfo {year} {1981})}\BibitemShut
  {NoStop}%
\bibitem [{\citenamefont {Sekino}\ and\ \citenamefont
  {Bartlett}(1984)}]{sekino1984}%
  \BibitemOpen
  \bibfield  {author} {\bibinfo {author} {\bibfnamefont {H.}~\bibnamefont
  {Sekino}}\ and\ \bibinfo {author} {\bibfnamefont {R.~J.}\ \bibnamefont
  {Bartlett}},\ }\bibfield  {title} {\enquote {\bibinfo {title} {A linear
  response, coupled-cluster theory for excitation energy},}\ }\href
  {https://doi.org/https://doi.org/10.1002/qua.560260826} {\bibfield  {journal}
  {\bibinfo  {journal} {Int. J. Quantum Chem.}\ }\textbf {\bibinfo {volume}
  {26}},\ \bibinfo {pages} {255--265} (\bibinfo {year} {1984})}\BibitemShut
  {NoStop}%
\bibitem [{\citenamefont {Comeau}\ and\ \citenamefont
  {Bartlett}(1993)}]{comeau1993}%
  \BibitemOpen
  \bibfield  {author} {\bibinfo {author} {\bibfnamefont {D.~C.}\ \bibnamefont
  {Comeau}}\ and\ \bibinfo {author} {\bibfnamefont {R.~J.}\ \bibnamefont
  {Bartlett}},\ }\href
  {https://doi.org/https://doi.org/10.1016/0009-2614(93)89023-B} {\bibfield
  {journal} {\bibinfo  {journal} {Chem. Phys. Lett.}\ }\textbf {\bibinfo
  {volume} {207}},\ \bibinfo {pages} {414--423} (\bibinfo {year}
  {1993})}\BibitemShut {NoStop}%
\bibitem [{\citenamefont {Stanton}\ and\ \citenamefont
  {Gauss}(1994)}]{stanton1994}%
  \BibitemOpen
  \bibfield  {author} {\bibinfo {author} {\bibfnamefont {J.~F.}\ \bibnamefont
  {Stanton}}\ and\ \bibinfo {author} {\bibfnamefont {J.}~\bibnamefont
  {Gauss}},\ }\bibfield  {title} {\enquote {\bibinfo {title} {Analytic energy
  derivatives for ionized states described by the equation‐of‐motion
  coupled cluster method},}\ }\href {https://doi.org/10.1063/1.468022}
  {\bibfield  {journal} {\bibinfo  {journal} {J. Chem. Phys.}\ }\textbf
  {\bibinfo {volume} {101}},\ \bibinfo {pages} {8938--8944} (\bibinfo {year}
  {1994})}\BibitemShut {NoStop}%
\bibitem [{\citenamefont {Bartlett}\ \emph {et~al.}(1997)\citenamefont
  {Bartlett}, \citenamefont {{Del Bene}}, \citenamefont {Perera},\ and\
  \citenamefont {Mattie}}]{bartlett1997}%
  \BibitemOpen
  \bibfield  {author} {\bibinfo {author} {\bibfnamefont {R.~J.}\ \bibnamefont
  {Bartlett}}, \bibinfo {author} {\bibfnamefont {J.~E.}\ \bibnamefont {{Del
  Bene}}}, \bibinfo {author} {\bibfnamefont {S.}~\bibnamefont {Perera}},\ and\
  \bibinfo {author} {\bibfnamefont {R.}~\bibnamefont {Mattie}},\ }\bibfield
  {title} {\enquote {\bibinfo {title} {Ammonia: the prototypical lone pair
  molecule},}\ }\href
  {https://doi.org/https://doi.org/10.1016/S0166-1280(97)90277-3} {\bibfield
  {journal} {\bibinfo  {journal} {J. of Mol.Struc.: THEOCHEM}\ }\textbf
  {\bibinfo {volume} {400}},\ \bibinfo {pages} {157--168} (\bibinfo {year}
  {1997})},\ \bibinfo {note} {ab Initio Benchmark Studies}\BibitemShut
  {NoStop}%
\bibitem [{\citenamefont {Stanton}\ and\ \citenamefont
  {Gauss}(1999)}]{stanton1999}%
  \BibitemOpen
  \bibfield  {author} {\bibinfo {author} {\bibfnamefont {J.~F.}\ \bibnamefont
  {Stanton}}\ and\ \bibinfo {author} {\bibfnamefont {J.}~\bibnamefont
  {Gauss}},\ }\bibfield  {title} {\enquote {\bibinfo {title} {A simple scheme
  for the direct calculation of ionization potentials with coupled-cluster
  theory that exploits established excitation energy methods},}\ }\href
  {https://doi.org/10.1063/1.479673} {\bibfield  {journal} {\bibinfo  {journal}
  {J. Chem. Phys.}\ }\textbf {\bibinfo {volume} {111}},\ \bibinfo {pages}
  {8785--8788} (\bibinfo {year} {1999})}\BibitemShut {NoStop}%
\bibitem [{\citenamefont {Kamiya}\ and\ \citenamefont
  {Hirata}(2006{\natexlab{b}})}]{muneaki2006}%
  \BibitemOpen
  \bibfield  {author} {\bibinfo {author} {\bibfnamefont {M.}~\bibnamefont
  {Kamiya}}\ and\ \bibinfo {author} {\bibfnamefont {S.}~\bibnamefont
  {Hirata}},\ }\bibfield  {title} {\enquote {\bibinfo {title} {Higher-order
  equation-of-motion coupled-cluster methods for ionization processes},}\
  }\href {https://doi.org/10.1063/1.2244570} {\bibfield  {journal} {\bibinfo
  {journal} {J. Chem. Phys.}\ }\textbf {\bibinfo {volume} {125}},\ \bibinfo
  {pages} {074111} (\bibinfo {year} {2006}{\natexlab{b}})}\BibitemShut
  {NoStop}%
\bibitem [{\citenamefont {Bomble}\ \emph {et~al.}(2005)\citenamefont {Bomble},
  \citenamefont {Saeh}, \citenamefont {Stanton}, \citenamefont {Szalay},
  \citenamefont {K{\'a}llay},\ and\ \citenamefont {Gauss}}]{bomble2005}%
  \BibitemOpen
  \bibfield  {author} {\bibinfo {author} {\bibfnamefont {Y.~J.}\ \bibnamefont
  {Bomble}}, \bibinfo {author} {\bibfnamefont {J.~C.}\ \bibnamefont {Saeh}},
  \bibinfo {author} {\bibfnamefont {J.~F.}\ \bibnamefont {Stanton}}, \bibinfo
  {author} {\bibfnamefont {P.~G.}\ \bibnamefont {Szalay}}, \bibinfo {author}
  {\bibfnamefont {M.}~\bibnamefont {K{\'a}llay}},\ and\ \bibinfo {author}
  {\bibfnamefont {J.}~\bibnamefont {Gauss}},\ }\bibfield  {title} {\enquote
  {\bibinfo {title} {Equation-of-motion coupled-cluster methods for ionized
  states with an approximate treatment of triple excitations},}\ }\href
  {https://doi.org/10.1063/1.1884600} {\bibfield  {journal} {\bibinfo
  {journal} {J. Chem. Phys.}\ }\textbf {\bibinfo {volume} {122}},\ \bibinfo
  {pages} {154107} (\bibinfo {year} {2005})}\BibitemShut {NoStop}%
\bibitem [{\citenamefont {Nooijen}\ and\ \citenamefont
  {Bartlett}(1995{\natexlab{a}})}]{nooijen1995a}%
  \BibitemOpen
  \bibfield  {author} {\bibinfo {author} {\bibfnamefont {M.}~\bibnamefont
  {Nooijen}}\ and\ \bibinfo {author} {\bibfnamefont {R.~J.}\ \bibnamefont
  {Bartlett}},\ }\bibfield  {title} {\enquote {\bibinfo {title} {Description of
  core‐excitation spectra by the open‐shell electron‐attachment
  equation‐of‐motion coupled cluster method},}\ }\href
  {https://doi.org/10.1063/1.469147} {\bibfield  {journal} {\bibinfo  {journal}
  {J. Chem. Phys.}\ }\textbf {\bibinfo {volume} {102}},\ \bibinfo {pages}
  {6735--6756} (\bibinfo {year} {1995}{\natexlab{a}})}\BibitemShut {NoStop}%
\bibitem [{\citenamefont {Nooijen}\ and\ \citenamefont
  {Bartlett}(1995{\natexlab{b}})}]{nooijen1995}%
  \BibitemOpen
  \bibfield  {author} {\bibinfo {author} {\bibfnamefont {M.}~\bibnamefont
  {Nooijen}}\ and\ \bibinfo {author} {\bibfnamefont {R.~J.}\ \bibnamefont
  {Bartlett}},\ }\bibfield  {title} {\enquote {\bibinfo {title} {Equation of
  motion coupled cluster method for electron attachment},}\ }\href
  {https://doi.org/10.1063/1.468592} {\bibfield  {journal} {\bibinfo  {journal}
  {J. Chem. Phys.}\ }\textbf {\bibinfo {volume} {102}},\ \bibinfo {pages}
  {3629--3647} (\bibinfo {year} {1995}{\natexlab{b}})}\BibitemShut {NoStop}%
\bibitem [{\citenamefont {Kamiya}\ and\ \citenamefont
  {Hirata}(2007)}]{muneaki2007}%
  \BibitemOpen
  \bibfield  {author} {\bibinfo {author} {\bibfnamefont {M.}~\bibnamefont
  {Kamiya}}\ and\ \bibinfo {author} {\bibfnamefont {S.}~\bibnamefont
  {Hirata}},\ }\bibfield  {title} {\enquote {\bibinfo {title} {Higher-order
  equation-of-motion coupled-cluster methods for electron attachment},}\ }\href
  {https://doi.org/10.1063/1.2715575} {\bibfield  {journal} {\bibinfo
  {journal} {J. Chem. Phys.}\ }\textbf {\bibinfo {volume} {126}},\ \bibinfo
  {pages} {134112} (\bibinfo {year} {2007})}\BibitemShut {NoStop}%
\bibitem [{\citenamefont {Hanrath}\ and\ \citenamefont
  {Engels-Putzka}(2010)}]{hanrath2010}%
  \BibitemOpen
  \bibfield  {author} {\bibinfo {author} {\bibfnamefont {M.}~\bibnamefont
  {Hanrath}}\ and\ \bibinfo {author} {\bibfnamefont {A.}~\bibnamefont
  {Engels-Putzka}},\ }\bibfield  {title} {\enquote {\bibinfo {title} {An
  efficient matrix-matrix multiplication based antisymmetric tensor contraction
  engine for general order coupled cluster},}\ }\href
  {https://doi.org/10.1063/1.3467878} {\bibfield  {journal} {\bibinfo
  {journal} {J. Chem. Phys.}\ }\textbf {\bibinfo {volume} {133}},\ \bibinfo
  {pages} {064108} (\bibinfo {year} {2010})}\BibitemShut {NoStop}%
\bibitem [{\citenamefont {Engels-Putzka}\ and\ \citenamefont
  {Hanrath}(2011)}]{engels2011}%
  \BibitemOpen
  \bibfield  {author} {\bibinfo {author} {\bibfnamefont {A.}~\bibnamefont
  {Engels-Putzka}}\ and\ \bibinfo {author} {\bibfnamefont {M.}~\bibnamefont
  {Hanrath}},\ }\bibfield  {title} {\enquote {\bibinfo {title} {A fully
  simultaneously optimizing genetic approach to the highly excited
  coupled-cluster factorization problem},}\ }\href
  {https://doi.org/10.1063/1.3561739} {\bibfield  {journal} {\bibinfo
  {journal} {J. Chem. Phys.}\ }\textbf {\bibinfo {volume} {134}},\ \bibinfo
  {pages} {124106} (\bibinfo {year} {2011})}\BibitemShut {NoStop}%
\bibitem [{\citenamefont {Kong}\ \emph {et~al.}(2009)\citenamefont {Kong},
  \citenamefont {Shamasundar}, \citenamefont {Demel},\ and\ \citenamefont
  {Nooijen}}]{kong2009}%
  \BibitemOpen
  \bibfield  {author} {\bibinfo {author} {\bibfnamefont {L.}~\bibnamefont
  {Kong}}, \bibinfo {author} {\bibfnamefont {K.~R.}\ \bibnamefont
  {Shamasundar}}, \bibinfo {author} {\bibfnamefont {O.}~\bibnamefont {Demel}},\
  and\ \bibinfo {author} {\bibfnamefont {M.}~\bibnamefont {Nooijen}},\
  }\bibfield  {title} {\enquote {\bibinfo {title} {State specific equation of
  motion coupled cluster method in general active space},}\ }\href
  {https://doi.org/10.1063/1.3089302} {\bibfield  {journal} {\bibinfo
  {journal} {J. Chem. Phys.}\ }\textbf {\bibinfo {volume} {130}},\ \bibinfo
  {pages} {114101} (\bibinfo {year} {2009})}\BibitemShut {NoStop}%
\bibitem [{\citenamefont {Das}, \citenamefont {K{\'a}llay},\ and\ \citenamefont
  {Mukherjee}(2010)}]{das2010}%
  \BibitemOpen
  \bibfield  {author} {\bibinfo {author} {\bibfnamefont {S.}~\bibnamefont
  {Das}}, \bibinfo {author} {\bibfnamefont {M.}~\bibnamefont {K{\'a}llay}},\
  and\ \bibinfo {author} {\bibfnamefont {D.}~\bibnamefont {Mukherjee}},\
  }\bibfield  {title} {\enquote {\bibinfo {title} {Inclusion of selected higher
  excitations involving active orbitals in the state-specific multireference
  coupled-cluster theory},}\ }\href {https://doi.org/10.1063/1.3515478}
  {\bibfield  {journal} {\bibinfo  {journal} {J. Chem. Phys.}\ }\textbf
  {\bibinfo {volume} {133}},\ \bibinfo {pages} {234110} (\bibinfo {year}
  {2010})}\BibitemShut {NoStop}%
\bibitem [{\citenamefont {Evangelista}\ and\ \citenamefont
  {Gauss}(2011)}]{evangelista2011}%
  \BibitemOpen
  \bibfield  {author} {\bibinfo {author} {\bibfnamefont {F.~A.}\ \bibnamefont
  {Evangelista}}\ and\ \bibinfo {author} {\bibfnamefont {J.}~\bibnamefont
  {Gauss}},\ }\bibfield  {title} {\enquote {\bibinfo {title} {An
  orbital-invariant internally contracted multireference coupled cluster
  approach},}\ }\href {https://doi.org/10.1063/1.3559149} {\bibfield  {journal}
  {\bibinfo  {journal} {J. Chem. Phys.}\ }\textbf {\bibinfo {volume} {134}},\
  \bibinfo {pages} {114102} (\bibinfo {year} {2011})}\BibitemShut {NoStop}%
\bibitem [{\citenamefont {Hanauer}\ and\ \citenamefont
  {K{\"o}hn}(2011)}]{hanauer2011}%
  \BibitemOpen
  \bibfield  {author} {\bibinfo {author} {\bibfnamefont {M.}~\bibnamefont
  {Hanauer}}\ and\ \bibinfo {author} {\bibfnamefont {A.}~\bibnamefont
  {K{\"o}hn}},\ }\bibfield  {title} {\enquote {\bibinfo {title} {Pilot
  applications of internally contracted multireference coupled cluster theory,
  and how to choose the cluster operator properly},}\ }\href
  {https://doi.org/10.1063/1.3592786} {\bibfield  {journal} {\bibinfo
  {journal} {J. Chem. Phys.}\ }\textbf {\bibinfo {volume} {134}},\ \bibinfo
  {pages} {204111} (\bibinfo {year} {2011})}\BibitemShut {NoStop}%
\bibitem [{\citenamefont {Hanauer}\ and\ \citenamefont
  {K{\"o}hn}(2012)}]{hanauer2012}%
  \BibitemOpen
  \bibfield  {author} {\bibinfo {author} {\bibfnamefont {M.}~\bibnamefont
  {Hanauer}}\ and\ \bibinfo {author} {\bibfnamefont {A.}~\bibnamefont
  {K{\"o}hn}},\ }\bibfield  {title} {\enquote {\bibinfo {title} {Communication:
  Restoring full size extensivity in internally contracted multireference
  coupled cluster theory},}\ }\href {https://doi.org/10.1063/1.4757728}
  {\bibfield  {journal} {\bibinfo  {journal} {J. Chem. Phys.}\ }\textbf
  {\bibinfo {volume} {137}},\ \bibinfo {pages} {131103} (\bibinfo {year}
  {2012})}\BibitemShut {NoStop}%
\bibitem [{\citenamefont {Shiozaki}\ \emph {et~al.}(2008)\citenamefont
  {Shiozaki}, \citenamefont {Kamiya}, \citenamefont {Hirata},\ and\
  \citenamefont {Valeev}}]{shiozaki2008}%
  \BibitemOpen
  \bibfield  {author} {\bibinfo {author} {\bibfnamefont {T.}~\bibnamefont
  {Shiozaki}}, \bibinfo {author} {\bibfnamefont {M.}~\bibnamefont {Kamiya}},
  \bibinfo {author} {\bibfnamefont {S.}~\bibnamefont {Hirata}},\ and\ \bibinfo
  {author} {\bibfnamefont {E.~F.}\ \bibnamefont {Valeev}},\ }\bibfield  {title}
  {\enquote {\bibinfo {title} {Equations of explicitly-correlated
  coupled-cluster methods},}\ }\href {https://doi.org/10.1039/B803704N}
  {\bibfield  {journal} {\bibinfo  {journal} {Phys. Chem. Chem. Phys.}\
  }\textbf {\bibinfo {volume} {10}},\ \bibinfo {pages} {3358--3370} (\bibinfo
  {year} {2008})}\BibitemShut {NoStop}%
\bibitem [{\citenamefont {K{\"o}hn}, \citenamefont {Richings},\ and\
  \citenamefont {Tew}(2008)}]{kohn2008}%
  \BibitemOpen
  \bibfield  {author} {\bibinfo {author} {\bibfnamefont {A.}~\bibnamefont
  {K{\"o}hn}}, \bibinfo {author} {\bibfnamefont {G.~W.}\ \bibnamefont
  {Richings}},\ and\ \bibinfo {author} {\bibfnamefont {D.~P.}\ \bibnamefont
  {Tew}},\ }\bibfield  {title} {\enquote {\bibinfo {title} {Implementation of
  the full explicitly correlated coupled-cluster singles and doubles model
  ccsd-f12 with optimally reduced auxiliary basis dependence},}\ }\href
  {https://doi.org/10.1063/1.3028546} {\bibfield  {journal} {\bibinfo
  {journal} {J. Chem. Phys.}\ }\textbf {\bibinfo {volume} {129}},\ \bibinfo
  {pages} {201103} (\bibinfo {year} {2008})}\BibitemShut {NoStop}%
\bibitem [{\citenamefont {MacLeod}\ and\ \citenamefont
  {Shiozaki}(2015)}]{matthew2015}%
  \BibitemOpen
  \bibfield  {author} {\bibinfo {author} {\bibfnamefont {M.~K.}\ \bibnamefont
  {MacLeod}}\ and\ \bibinfo {author} {\bibfnamefont {T.}~\bibnamefont
  {Shiozaki}},\ }\bibfield  {title} {\enquote {\bibinfo {title} {Communication:
  Automatic code generation enables nuclear gradient computations for fully
  internally contracted multireference theory},}\ }\href
  {https://doi.org/10.1063/1.4907717} {\bibfield  {journal} {\bibinfo
  {journal} {J. Chem. Phys.}\ }\textbf {\bibinfo {volume} {142}},\ \bibinfo
  {pages} {051103} (\bibinfo {year} {2015})}\BibitemShut {NoStop}%
\bibitem [{\citenamefont {Rubin}\ and\ \citenamefont {III}(2021)}]{rubin2021}%
  \BibitemOpen
  \bibfield  {author} {\bibinfo {author} {\bibfnamefont {N.~C.}\ \bibnamefont
  {Rubin}}\ and\ \bibinfo {author} {\bibfnamefont {A.~E.~D.}\ \bibnamefont
  {III}},\ }\bibfield  {title} {\enquote {\bibinfo {title} {p†q: a tool for
  prototyping many-body methods for quantum chemistry},}\ }\href
  {https://doi.org/10.1080/00268976.2021.1954709} {\bibfield  {journal}
  {\bibinfo  {journal} {Mol. Phys.}\ }\textbf {\bibinfo {volume} {119}},\
  \bibinfo {pages} {e1954709} (\bibinfo {year} {2021})}\BibitemShut {NoStop}%
\bibitem [{\citenamefont {Evangelista}(2022)}]{evangelista2022}%
  \BibitemOpen
  \bibfield  {author} {\bibinfo {author} {\bibfnamefont {F.~A.}\ \bibnamefont
  {Evangelista}},\ }\bibfield  {title} {\enquote {\bibinfo {title} {Automatic
  derivation of many-body theories based on general fermi vacua},}\ }\href
  {https://doi.org/10.1063/5.0097858} {\bibfield  {journal} {\bibinfo
  {journal} {J. Chem. Phys.}\ }\textbf {\bibinfo {volume} {157}},\ \bibinfo
  {pages} {064111} (\bibinfo {year} {2022})}\BibitemShut {NoStop}%
\bibitem [{\citenamefont {Mukherjee}(1997)}]{mukherjee1997}%
  \BibitemOpen
  \bibfield  {author} {\bibinfo {author} {\bibfnamefont {D.}~\bibnamefont
  {Mukherjee}},\ }\bibfield  {title} {\enquote {\bibinfo {title} {Normal
  ordering and a wick-like reduction theorem for fermions with respect to a
  multi-determinantal reference state},}\ }\href
  {https://doi.org/https://doi.org/10.1016/S0009-2614(97)00714-8} {\bibfield
  {journal} {\bibinfo  {journal} {Chem. Phys. Lett.}\ }\textbf {\bibinfo
  {volume} {274}},\ \bibinfo {pages} {561--566} (\bibinfo {year}
  {1997})}\BibitemShut {NoStop}%
\bibitem [{\citenamefont {Kutzelnigg}\ and\ \citenamefont
  {Mukherjee}(1997)}]{kutzelnigg1997}%
  \BibitemOpen
  \bibfield  {author} {\bibinfo {author} {\bibfnamefont {W.}~\bibnamefont
  {Kutzelnigg}}\ and\ \bibinfo {author} {\bibfnamefont {D.}~\bibnamefont
  {Mukherjee}},\ }\bibfield  {title} {\enquote {\bibinfo {title} {Normal order
  and extended wick theorem for a multiconfiguration reference wave
  function},}\ }\href {https://doi.org/10.1063/1.474405} {\bibfield  {journal}
  {\bibinfo  {journal} {J. Chem. Phys.}\ }\textbf {\bibinfo {volume} {107}},\
  \bibinfo {pages} {432--449} (\bibinfo {year} {1997})}\BibitemShut {NoStop}%
\bibitem [{\citenamefont {Monino}\ and\ \citenamefont
  {Loos}(2022)}]{monino2022}%
  \BibitemOpen
  \bibfield  {author} {\bibinfo {author} {\bibfnamefont {E.}~\bibnamefont
  {Monino}}\ and\ \bibinfo {author} {\bibfnamefont {P.-F.}\ \bibnamefont
  {Loos}},\ }\bibfield  {title} {\enquote {\bibinfo {title} {Unphysical
  discontinuities, intruder states and regularization in gw methods},}\ }\href
  {https://doi.org/10.1063/5.0089317} {\bibfield  {journal} {\bibinfo
  {journal} {J. Chem. Phys.}\ }\textbf {\bibinfo {volume} {156}},\ \bibinfo
  {pages} {231101} (\bibinfo {year} {2022})}\BibitemShut {NoStop}%
\bibitem [{\citenamefont {Berkelbach}(2018)}]{berkelbach2018}%
  \BibitemOpen
  \bibfield  {author} {\bibinfo {author} {\bibfnamefont {T.~C.}\ \bibnamefont
  {Berkelbach}},\ }\bibfield  {title} {\enquote {\bibinfo {title}
  {Communication: Random-phase approximation excitation energies from
  approximate equation-of-motion coupled-cluster doubles},}\ }\href
  {https://doi.org/10.1063/1.5032314} {\bibfield  {journal} {\bibinfo
  {journal} {J. Chem. Phys.}\ }\textbf {\bibinfo {volume} {149}},\ \bibinfo
  {pages} {041103} (\bibinfo {year} {2018})}\BibitemShut {NoStop}%
\bibitem [{\citenamefont {Scuseria}, \citenamefont {Henderson},\ and\
  \citenamefont {Sorensen}(2008)}]{scuseria2008}%
  \BibitemOpen
  \bibfield  {author} {\bibinfo {author} {\bibfnamefont {G.~E.}\ \bibnamefont
  {Scuseria}}, \bibinfo {author} {\bibfnamefont {T.~M.}\ \bibnamefont
  {Henderson}},\ and\ \bibinfo {author} {\bibfnamefont {D.~C.}\ \bibnamefont
  {Sorensen}},\ }\bibfield  {title} {\enquote {\bibinfo {title} {The ground
  state correlation energy of the random phase approximation from a ring
  coupled cluster doubles approach},}\ }\href
  {https://doi.org/10.1063/1.3043729} {\bibfield  {journal} {\bibinfo
  {journal} {J. Chem. Phys.}\ }\textbf {\bibinfo {volume} {129}},\ \bibinfo
  {pages} {231101} (\bibinfo {year} {2008})}\BibitemShut {NoStop}%
\bibitem [{\citenamefont {Scuseria}, \citenamefont {Henderson},\ and\
  \citenamefont {Bulik}(2013)}]{scuseria2013}%
  \BibitemOpen
  \bibfield  {author} {\bibinfo {author} {\bibfnamefont {G.~E.}\ \bibnamefont
  {Scuseria}}, \bibinfo {author} {\bibfnamefont {T.~M.}\ \bibnamefont
  {Henderson}},\ and\ \bibinfo {author} {\bibfnamefont {I.~W.}\ \bibnamefont
  {Bulik}},\ }\bibfield  {title} {\enquote {\bibinfo {title} {Particle-particle
  and quasiparticle random phase approximations: Connections to coupled cluster
  theory},}\ }\href {https://doi.org/10.1063/1.4820557} {\bibfield  {journal}
  {\bibinfo  {journal} {J. Chem. Phys.}\ }\textbf {\bibinfo {volume} {139}},\
  \bibinfo {pages} {104113} (\bibinfo {year} {2013})}\BibitemShut {NoStop}%
\bibitem [{\citenamefont {Lange}\ and\ \citenamefont
  {Berkelbach}(2018)}]{lange2018}%
  \BibitemOpen
  \bibfield  {author} {\bibinfo {author} {\bibfnamefont {M.~F.}\ \bibnamefont
  {Lange}}\ and\ \bibinfo {author} {\bibfnamefont {T.~C.}\ \bibnamefont
  {Berkelbach}},\ }\bibfield  {title} {\enquote {\bibinfo {title} {On the
  relation between equation-of-motion coupled-cluster theory and the gw
  approximation},}\ }\href {https://doi.org/10.1021/acs.jctc.8b00455}
  {\bibfield  {journal} {\bibinfo  {journal} {J. Chem. Theory Comput.}\
  }\textbf {\bibinfo {volume} {14}},\ \bibinfo {pages} {4224--4236} (\bibinfo
  {year} {2018})},\ \bibinfo {note} {pMID: 30028614}\BibitemShut {NoStop}%
\bibitem [{\citenamefont {Inc.}()}]{mathematica}%
  \BibitemOpen
  \bibfield  {author} {\bibinfo {author} {\bibfnamefont {W.~R.}\ \bibnamefont
  {Inc.}},\ }\href {https://www.wolfram.com/mathematica} {\enquote {\bibinfo
  {title} {Mathematica, {V}ersion 13.2},}\ }\bibinfo {note} {Champaign, IL,
  2022}\BibitemShut {NoStop}%
\bibitem [{\citenamefont {Helgaker}, \citenamefont {J{\o}rgensen},\ and\
  \citenamefont {Olsen}(2000)}]{helgaker2013}%
  \BibitemOpen
  \bibfield  {author} {\bibinfo {author} {\bibfnamefont {T.}~\bibnamefont
  {Helgaker}}, \bibinfo {author} {\bibfnamefont {P.}~\bibnamefont
  {J{\o}rgensen}},\ and\ \bibinfo {author} {\bibfnamefont {J.}~\bibnamefont
  {Olsen}},\ }\href {https://doi.org/10.1002/9781119019572} {\emph {\bibinfo
  {title} {Molecular Electronic Structure Theory}}}\ (\bibinfo  {publisher}
  {John Wiley \& Sons, LTD},\ \bibinfo {address} {Chichester},\ \bibinfo {year}
  {2000})\BibitemShut {NoStop}%
\bibitem [{\citenamefont {Surj{\'a}n}(2012)}]{surjan2012}%
  \BibitemOpen
  \bibfield  {author} {\bibinfo {author} {\bibfnamefont {P.~R.}\ \bibnamefont
  {Surj{\'a}n}},\ }\href
  {https://doi.org/https://doi.org/10.1007/978-3-642-74755-7} {\emph {\bibinfo
  {title} {Second quantized approach to quantum chemistry: an elementary
  introduction}}}\ (\bibinfo  {publisher} {Springer Science \& Business
  Media},\ \bibinfo {year} {2012})\BibitemShut {NoStop}%
\bibitem [{\citenamefont {Szabo}\ and\ \citenamefont
  {Ostlund}(2012)}]{szabo2012}%
  \BibitemOpen
  \bibfield  {author} {\bibinfo {author} {\bibfnamefont {A.}~\bibnamefont
  {Szabo}}\ and\ \bibinfo {author} {\bibfnamefont {N.~S.}\ \bibnamefont
  {Ostlund}},\ }\href@noop {} {\emph {\bibinfo {title} {Modern Quantum
  Chemistry: Introduction to Advanced Electronic Structure Theory}}}\ (\bibinfo
   {publisher} {Courier Corporation},\ \bibinfo {year} {2012})\BibitemShut
  {NoStop}%
\bibitem [{\citenamefont {Bogoliubov}, \citenamefont {Shirkov},\ and\
  \citenamefont {Henley}(1960)}]{bogoliubob1959}%
  \BibitemOpen
  \bibfield  {author} {\bibinfo {author} {\bibfnamefont {N.~N.}\ \bibnamefont
  {Bogoliubov}}, \bibinfo {author} {\bibfnamefont {D.~V.}\ \bibnamefont
  {Shirkov}},\ and\ \bibinfo {author} {\bibfnamefont {E.~M.}\ \bibnamefont
  {Henley}},\ }\href@noop {} {\emph {\bibinfo {title} {{Introduction to the
  Theory of Quantized Fields}}}}\ (\bibinfo {year} {1960})\BibitemShut
  {NoStop}%
\bibitem [{\citenamefont {Paldus}\ and\ \citenamefont {{\v C}{\'\i}{\v
  z}ek}(1975)}]{paldus1975}%
  \BibitemOpen
  \bibfield  {author} {\bibinfo {author} {\bibfnamefont {J.}~\bibnamefont
  {Paldus}}\ and\ \bibinfo {author} {\bibfnamefont {J.}~\bibnamefont {{\v
  C}{\'\i}{\v z}ek}},\ }\bibfield  {title} {\enquote {\bibinfo {title}
  {Time-independent diagrammatic approach to perturbation theory of fermion
  systems},}\ \ }(\bibinfo  {publisher} {Academic Press},\ \bibinfo {year}
  {1975})\ pp.\ \bibinfo {pages} {105--197}\BibitemShut {NoStop}%
\bibitem [{\citenamefont {Merzbacher}(1998)}]{merzbacher1998}%
  \BibitemOpen
  \bibfield  {author} {\bibinfo {author} {\bibfnamefont {E.}~\bibnamefont
  {Merzbacher}},\ }\href@noop {} {\emph {\bibinfo {title} {{Quantum
  Mechanics}}}}\ (\bibinfo  {publisher} {John Wiley \& Sons},\ \bibinfo {year}
  {1998})\BibitemShut {NoStop}%
\bibitem [{\citenamefont {Hubbard}\ and\ \citenamefont
  {Peierls}(1957)}]{hubbard1957}%
  \BibitemOpen
  \bibfield  {author} {\bibinfo {author} {\bibfnamefont {J.}~\bibnamefont
  {Hubbard}}\ and\ \bibinfo {author} {\bibfnamefont {R.~E.}\ \bibnamefont
  {Peierls}},\ }\bibfield  {title} {\enquote {\bibinfo {title} {The description
  of collective motions in terms of many-body perturbation theory},}\ }\href
  {https://doi.org/10.1098/rspa.1957.0106} {\bibfield  {journal} {\bibinfo
  {journal} {Proc. R. Soc. London.}\ }\textbf {\bibinfo {volume} {240}},\
  \bibinfo {pages} {539--560} (\bibinfo {year} {1957})}\BibitemShut {NoStop}%
\bibitem [{\citenamefont {Hubbard}\ and\ \citenamefont
  {Peierls}(1958)}]{hubbard1958}%
  \BibitemOpen
  \bibfield  {author} {\bibinfo {author} {\bibfnamefont {J.}~\bibnamefont
  {Hubbard}}\ and\ \bibinfo {author} {\bibfnamefont {R.~E.}\ \bibnamefont
  {Peierls}},\ }\bibfield  {title} {\enquote {\bibinfo {title} {The description
  of collective motions in terms of many-body perturbation theory. ii. the
  correlation energy of a free-electron gas},}\ }\href
  {https://doi.org/10.1098/rspa.1958.0003} {\bibfield  {journal} {\bibinfo
  {journal} {Proc. R. Soc. London.}\ }\textbf {\bibinfo {volume} {243}},\
  \bibinfo {pages} {336--352} (\bibinfo {year} {1958})}\BibitemShut {NoStop}%
\bibitem [{\citenamefont {Jensen}(2017)}]{jensen2017}%
  \BibitemOpen
  \bibfield  {author} {\bibinfo {author} {\bibfnamefont {F.}~\bibnamefont
  {Jensen}},\ }\href@noop {} {\emph {\bibinfo {title} {{Introduction to
  Computational Chemistry}}}}\ (\bibinfo  {publisher} {John wiley and sons},\
  \bibinfo {year} {2017})\BibitemShut {NoStop}%
\bibitem [{\citenamefont {Kumar}(1965)}]{kumar1965}%
  \BibitemOpen
  \bibfield  {author} {\bibinfo {author} {\bibfnamefont {K.}~\bibnamefont
  {Kumar}},\ }\bibfield  {title} {\enquote {\bibinfo {title} {On expanding the
  exponential},}\ }\href {https://doi.org/10.1063/1.1704742} {\bibfield
  {journal} {\bibinfo  {journal} {J. Math. Phys.}\ }\textbf {\bibinfo {volume}
  {6}},\ \bibinfo {pages} {1928--1934} (\bibinfo {year} {1965})}\BibitemShut
  {NoStop}%
\bibitem [{\citenamefont {Sattelmeyer}, \citenamefont {{Schaefer III}},\ and\
  \citenamefont {Stanton}(2003)}]{sattelmeyer2003}%
  \BibitemOpen
  \bibfield  {author} {\bibinfo {author} {\bibfnamefont {K.~W.}\ \bibnamefont
  {Sattelmeyer}}, \bibinfo {author} {\bibfnamefont {H.~F.}\ \bibnamefont
  {{Schaefer III}}},\ and\ \bibinfo {author} {\bibfnamefont {J.~F.}\
  \bibnamefont {Stanton}},\ }\bibfield  {title} {\enquote {\bibinfo {title}
  {Use of 2h and 3h−p-like coupled-cluster tamm--dancoff approaches for the
  equilibrium properties of ozone},}\ }\href
  {https://doi.org/https://doi.org/10.1016/S0009-2614(03)01181-3} {\bibfield
  {journal} {\bibinfo  {journal} {Chem. Phys. Lett.}\ }\textbf {\bibinfo
  {volume} {378}},\ \bibinfo {pages} {42--46} (\bibinfo {year}
  {2003})}\BibitemShut {NoStop}%
\bibitem [{\citenamefont {Demel}\ \emph {et~al.}(2008)\citenamefont {Demel},
  \citenamefont {Shamasundar}, \citenamefont {Kong},\ and\ \citenamefont
  {Nooijen}}]{demel2008}%
  \BibitemOpen
  \bibfield  {author} {\bibinfo {author} {\bibfnamefont {O.}~\bibnamefont
  {Demel}}, \bibinfo {author} {\bibfnamefont {K.~R.}\ \bibnamefont
  {Shamasundar}}, \bibinfo {author} {\bibfnamefont {L.}~\bibnamefont {Kong}},\
  and\ \bibinfo {author} {\bibfnamefont {M.}~\bibnamefont {Nooijen}},\
  }\bibfield  {title} {\enquote {\bibinfo {title} {Application of double
  ionization state-specific equation of motion coupled cluster method to
  organic diradicals},}\ }\href {https://doi.org/10.1021/jp800577q} {\bibfield
  {journal} {\bibinfo  {journal} {J. Phys. Chem. A}\ }\textbf {\bibinfo
  {volume} {112}},\ \bibinfo {pages} {11895--11902} (\bibinfo {year}
  {2008})}\BibitemShut {NoStop}%
\bibitem [{\citenamefont {Musia{\l}}, \citenamefont {Perera},\ and\
  \citenamefont {Bartlett}(2011)}]{musial2011}%
  \BibitemOpen
  \bibfield  {author} {\bibinfo {author} {\bibfnamefont {M.}~\bibnamefont
  {Musia{\l}}}, \bibinfo {author} {\bibfnamefont {A.}~\bibnamefont {Perera}},\
  and\ \bibinfo {author} {\bibfnamefont {R.~J.}\ \bibnamefont {Bartlett}},\
  }\bibfield  {title} {\enquote {\bibinfo {title} {Multireference
  coupled-cluster theory: The easy way},}\ }\href
  {https://doi.org/10.1063/1.3567115} {\bibfield  {journal} {\bibinfo
  {journal} {J. Chem. Phys.}\ }\textbf {\bibinfo {volume} {134}},\ \bibinfo
  {pages} {114108} (\bibinfo {year} {2011})}\BibitemShut {NoStop}%
\bibitem [{\citenamefont {Ku{\'s}}\ and\ \citenamefont
  {Krylov}(2011)}]{kus2011}%
  \BibitemOpen
  \bibfield  {author} {\bibinfo {author} {\bibfnamefont {T.}~\bibnamefont
  {Ku{\'s}}}\ and\ \bibinfo {author} {\bibfnamefont {A.~I.}\ \bibnamefont
  {Krylov}},\ }\bibfield  {title} {\enquote {\bibinfo {title} {Using the
  charge-stabilization technique in the double ionization potential
  equation-of-motion calculations with dianion references},}\ }\href
  {https://doi.org/10.1063/1.3626149} {\bibfield  {journal} {\bibinfo
  {journal} {J. Chem. Phys.}\ }\textbf {\bibinfo {volume} {135}},\ \bibinfo
  {pages} {084109} (\bibinfo {year} {2011})}\BibitemShut {NoStop}%
\bibitem [{\citenamefont {Shen}\ and\ \citenamefont
  {Piecuch}(2013)}]{shen2013}%
  \BibitemOpen
  \bibfield  {author} {\bibinfo {author} {\bibfnamefont {J.}~\bibnamefont
  {Shen}}\ and\ \bibinfo {author} {\bibfnamefont {P.}~\bibnamefont {Piecuch}},\
  }\bibfield  {title} {\enquote {\bibinfo {title} {Doubly electron-attached and
  doubly ionized equation-of-motion coupled-cluster methods with
  4-particle--2-hole and 4-hole--2-particle excitations and their active-space
  extensions},}\ }\href {https://doi.org/10.1063/1.4803883} {\bibfield
  {journal} {\bibinfo  {journal} {J. Chem. Phys.}\ }\textbf {\bibinfo {volume}
  {138}},\ \bibinfo {pages} {194102} (\bibinfo {year} {2013})}\BibitemShut
  {NoStop}%
\bibitem [{\citenamefont {Kucharski}\ and\ \citenamefont
  {Bartlett}(1991)}]{kucharski1991}%
  \BibitemOpen
  \bibfield  {author} {\bibinfo {author} {\bibfnamefont {S.~A.}\ \bibnamefont
  {Kucharski}}\ and\ \bibinfo {author} {\bibfnamefont {R.~J.}\ \bibnamefont
  {Bartlett}},\ }\bibfield  {title} {\enquote {\bibinfo {title} {Recursive
  intermediate factorization and complete computational linearization of the
  coupled-cluster single, double, triple, and quadruple excitation
  equations},}\ }\href {https://doi.org/https://doi.org/10.1007/BF01117419}
  {\bibfield  {journal} {\bibinfo  {journal} {Theoret. Chim. Acta}\ }\textbf
  {\bibinfo {volume} {80}},\ \bibinfo {pages} {387--405} (\bibinfo {year}
  {1991})}\BibitemShut {NoStop}%
\bibitem [{\citenamefont {Kucharski}\ and\ \citenamefont
  {Bartlett}(1992)}]{kucharski1992}%
  \BibitemOpen
  \bibfield  {author} {\bibinfo {author} {\bibfnamefont {S.~A.}\ \bibnamefont
  {Kucharski}}\ and\ \bibinfo {author} {\bibfnamefont {R.~J.}\ \bibnamefont
  {Bartlett}},\ }\bibfield  {title} {\enquote {\bibinfo {title} {The
  coupled-cluster single, double, triple, and quadruple excitation method},}\
  }\href {https://doi.org/https://doi.org/10.1063/1.463930} {\bibfield
  {journal} {\bibinfo  {journal} {J. Chem. Phys.}\ }\textbf {\bibinfo {volume}
  {97}},\ \bibinfo {pages} {4282--4288} (\bibinfo {year} {1992})}\BibitemShut
  {NoStop}%
\bibitem [{\citenamefont {Musia{\l}}, \citenamefont {Kucharski},\ and\
  \citenamefont {Bartlett}(2002{\natexlab{a}})}]{musial2002}%
  \BibitemOpen
  \bibfield  {author} {\bibinfo {author} {\bibfnamefont {M.}~\bibnamefont
  {Musia{\l}}}, \bibinfo {author} {\bibfnamefont {S.}~\bibnamefont
  {Kucharski}},\ and\ \bibinfo {author} {\bibfnamefont {R.}~\bibnamefont
  {Bartlett}},\ }\href {https://doi.org/https://doi.org/10.1063/1.1445744}
  {\bibfield  {journal} {\bibinfo  {journal} {J. Chem. Phys.}\ }\textbf
  {\bibinfo {volume} {116}},\ \bibinfo {pages} {4382--4388} (\bibinfo {year}
  {2002}{\natexlab{a}})}\BibitemShut {NoStop}%
\bibitem [{\citenamefont {Musia{\l}}, \citenamefont {Kucharski},\ and\
  \citenamefont {Bartlett}(2002{\natexlab{b}})}]{musial2002a}%
  \BibitemOpen
  \bibfield  {author} {\bibinfo {author} {\bibfnamefont {M.}~\bibnamefont
  {Musia{\l}}}, \bibinfo {author} {\bibfnamefont {S.}~\bibnamefont
  {Kucharski}},\ and\ \bibinfo {author} {\bibfnamefont {R.}~\bibnamefont
  {Bartlett}},\ }\href
  {https://doi.org/https://doi.org/10.1080/00268970110120319} {\bibfield
  {journal} {\bibinfo  {journal} {Mol. Phys.}\ }\textbf {\bibinfo {volume}
  {100}},\ \bibinfo {pages} {1867--1872} (\bibinfo {year}
  {2002}{\natexlab{b}})}\BibitemShut {NoStop}%
\bibitem [{\citenamefont {Davidson}(1975)}]{davidson1975}%
  \BibitemOpen
  \bibfield  {author} {\bibinfo {author} {\bibfnamefont {E.~R.}\ \bibnamefont
  {Davidson}},\ }\bibfield  {title} {\enquote {\bibinfo {title} {The iterative
  calculation of a few of the lowest eigenvalues and corresponding eigenvectors
  of large real-symmetric matrices},}\ }\href
  {https://doi.org/10.1016/0021-9991(75)90065-0} {\bibfield  {journal}
  {\bibinfo  {journal} {J. Comput. Phys.}\ }\textbf {\bibinfo {volume} {17}},\
  \bibinfo {pages} {87--94} (\bibinfo {year} {1975})}\BibitemShut {NoStop}%
\bibitem [{\citenamefont {Quintero-Monsebaiz}\ and\ \citenamefont
  {Loos}(2023)}]{rauleomccgen2023}%
  \BibitemOpen
  \bibfield  {author} {\bibinfo {author} {\bibfnamefont {R.}~\bibnamefont
  {Quintero-Monsebaiz}}\ and\ \bibinfo {author} {\bibfnamefont {P.-F.}\
  \bibnamefont {Loos}},\ }\href {https://github.com/rquintero-88/eomccgen.git}
  {\enquote {\bibinfo {title} {eomccgen},}\ } (\bibinfo {year}
  {2023})\BibitemShut {NoStop}%
\end{thebibliography}%
\end{document}